\documentclass[a4wide]{svjour3}       
\usepackage[a4paper,bindingoffset=0.2in,%
            left=1in,right=1in,top=1in,bottom=1in,%
            footskip=.25in]{geometry}
\usepackage{blindtext}
\RequirePackage{fix-cm}
\usepackage{a4wide}

\smartqed  
\usepackage{graphicx,inputenc}
\newcommand{\mysubsection}[1]{\subsection{\colorbox{Gray}}}
\def\atmosfair{{\it atmosfair}}
\def\myclimate{{\it myclimate}}
\def\climatecare{{\it climatecare}}
\def\defra{{\it DEFRA}}
\def\ademe{{\it ADEME}}
\def\icao{{\it ICAO}}
\def\co2eq{CO$_2{\rm eq}$}
\def\co2{CO$_2$}
\usepackage[colorlinks=true,linkcolor=blue,citecolor=red,allcolors=red]{hyperref}%
\usepackage{amsmath}
\begin{document}

\title{Estimating, monitoring and minimizing the travel footprint associated with the development of the Athena X-ray Integral Field Unit} 
\subtitle{An on-line travel footprint calculator released to the science community}


\author{Didier Barret} 
\institute{CNRS, Institut de Recherche en Astrophysique et Plan\'etologie, 9 Avenue du colonel Roche, BP 44346, F-31028 Toulouse Cedex 4, France \email{dbarret@irap.omp.eu}}

\titlerunning{Monitoring and reducing the travel footprint of large space projects} 
\date{Accepted for publication on April 8th, 2020. Published in June 2020 in Experimental Astronomy, Volume 49, Issue 3, p.183-216, \url{https://ui.adsabs.harvard.edu/abs/2020ExA....49..183B/abstract}} 

\maketitle 



\begin{abstract}
Global warming imposes us to reflect on the way we carry research, embarking on the obligation to minimize the environmental impact of our research programs, with the reduction of our travel footprint being one of the easiest actions to implement, thanks to the advance of digital technology. The X-ray Integral Field Unit (X-IFU), the cryogenic spectrometer of the Athena space X-ray observatory of the European Space Agency will be developed by a large international consortium, currently involving $\sim 240$ members, split over 13 countries, 11 in Europe, Japan and the United States. The travel footprint associated with the development of the X-IFU is to be minimized. For that purpose, a travel footprint calculator has been developed and released to the X-IFU consortium members. The calculator uses seven different emission factors and methods leading to estimates that differ by up to a factor of $\sim 5$ for the same flying distance. These differences illustrate the lack of standards and regulations for computing the footprint of flight travels and are explained primarily, though partly, by different accounting of non-\co2\ effects. When accounting for non-\co2\ effects, the flight emission is estimated as a multiple of the direct \co2\ emission from burning fuel, expressed in \co2-equivalent (\co2eq), with a multiplication factor ranging from 2 to 3. Considering or ignoring this multiplication factor is key when comparing alternative modes of transportation to flying. The calculator enables us to compute the travel footprint of a large set of travels and can help identify a meeting place that minimizes the overall travel footprint for a large set of possible city hosts, e.g. cities with large airports. The calculator also includes the option for a minimum distance above which flying is considered the most suitable transport option ; below that chosen distance, the emission of train journeys are considered. To demonstrate its full capabilities, the calculator is first run on one of the largest scientific meetings; the fall meeting of the American Geoscience Union (AGU) gathering some $\sim 24000$ participants and the four meetings of the lead authors of the working group I of the  Intergovernmental Panel on Climate Change (IPCC) preparing its sixth assessment report. In both examples, the calculator is used to compute the location of the meetings that would minimize the travel footprint. Then, the travel footprint of a representative set of X-IFU related meetings is estimated to be $\sim 500$ tons of \co2eq\ per year (to place this number in perspective, it is equivalent to $\sim 2$ million kilometers driven by an average passenger vehicle). Of this amount, each annual consortium meeting accounts for $\sim 100$ tons, being located at a site of minimum emission and for a minimum distance for flying of 700 km. Actions to reduce the X-IFU travel footprint are being implemented, e.g., the number of large consortium meetings has been reduced to one per year and face-to-face working meetings are progressively replaced by video conferences. As the on-line travel footprint calculator may be used to all scientific collaborations and meetings, the calculator and its methodology described in this paper are made freely available to the science community\footnote{\url{https://travel-footprint-calculator.irap.omp.eu}}.
\end{abstract}


\section{Introduction} Global warming poses a threat for the habitability of our planet, calling for rapid actions from all sectors to reduce \co2\ emissions, as well as the net effect of non-\co2\ emissions. More specifically, in order to keep global warning below $1.5^\circ$C, \co2\ emission must be reduced globally by 45\% by 2030 (from the 2010 values) with the need to reach net zero by 2050 \cite{ipccreport2019}. In 2018, the total \co2\ emissions from all aviation services was 918 million metric tons. That was 2.4\% of global \co2\ emissions from fossil fuel use, with a 32\% increase over the last five years \cite{graver}. Aviation contributes to $\sim 10$\% of all Greenhouse Gas (GHG) emissions from all transportation sources \cite{sims}. The demand for flying is expected to increase following the growth and development of the world economy, the development of trade and commerce and cultural exchanges among peoples and nations, thus leading to an increase of the aircraft emission.  At the current pace, aircraft emissions of carbon dioxide may reach $\sim 2400$ million metric tons by 2050, with an annual traffic growth rate of 4.6\% and a 2\% efficiency saving, e.g. \cite{terrenoire,tabuchi} and references therein. 

In addition, it is generally agreed that aviation contributes to climate change more than just with the emission of \co2\ from burning fuels, by releasing gases and particles directly into the upper troposphere and lower stratosphere where they have an impact on atmospheric composition. Atmospheric changes from aircraft result from three types of processes: direct emission of radiatively active substances (e.g., \co2\ or water vapor); emission of chemical species that produce or destroy radiatively active substances (e.g., NOx, which modifies O$_3$, CH$_4$ concentrations); and emission of substances that trigger the generation of aerosol particles or lead to changes in natural clouds (e.g., contrails) \cite{ipcc99,lee2009,shine}. While the radiative forcing due to changes in \co2\ is as well characterized as those from any other sources due to human activity,  estimating the forcings of non-\co2\ agents acting on shorter timescales than \co2\ is much more challenging and still subject to some uncertainties \cite{shine}, see also \cite{terrenoire}. Because of that, in the tools commonly used to compute the emissions of flights, either non-\co2\ effects are not accounted for, or simply modeled by a multiplier of the direct \co2\ emission. The \co2-multiplier varies between 1.3 to 2.0, based on the estimates of the Global Warming Potential\footnote{Global Warming Potential (GWP): An index, based on radiative properties of greenhouse gases, measuring the radiative forcing following a pulse emission of a unit mass of a given greenhouse gas in the present-day atmosphere integrated over a chosen time horizon, relative to that of carbon dioxide. The GWP represents the combined effect of the differing times these gases remain in the atmosphere and their relative effectiveness in causing radiative forcing. The Kyoto Protocol is based on GWPs from pulse emissions over a 100-year time frame.} of aviation emissions integrated over a 100 year timescale \cite{shine}. 




In response to the criticality of the situation, the International Civil Aviation Organization (\icao), as a specialized United Nations agency to address all matters related to international civil aviation, including environmental protection, and its member states have committed together to the so-called {\it Carbon neutral growth from 2020} onwards. They define what they call as {\it aspirational} goal of keeping the global net \co2\ emissions from international aviation from 2020 at the same level, through technological and operational improvements, the use of sustainable aviation fuels and the Carbon Offsetting and Reduction Scheme for International Aviation (CORSIA)\footnote{\url{https://www.icao.int/environmental-protection/CORSIA/Documents/CORSIA_FAQs_October\%202019_final.pdf}}. The above approach has however received a fair amount of criticisms and its overall effectiveness would need to be demonstrated, e.g. the controversial biofuels, especially those from palm oil that would result in more emissions even than the fossil fuels they replace and are a cause of biodiversity loss, deforestation and human rights abuses, or because offsetting diverts the focus from reducing emissions to trading on emissions\footnote{Criticisms on the CORSIA process are briefly listed at the URL \url{https://en.wikipedia.org/wiki/Carbon_Offsetting_and_Reduction_Scheme_for_International_Aviation} and references therein., see also \cite{warnecke} showing that the CORSIA scheme will only compensate for the emissions increase if robust criteria for the eligibility of offset credits are adopted.}. This has been re-emphasized in the IPCC 2019 report on climate change and land (SRCCL) which flagged the limited potential to use additional land for afforestation and bioenergy production, in a context of growing concerns for increasing pressures on land, conservation of terrestrial biodiversity and food security \cite{skula}.

Because carrying research is generally associated with traveling all across the world (especially since the 80s), our GHG emission is for a vast majority of scientists, most notably for the most senior ones, e.g. \cite{epfl}, dominated by travels. ETH Zurich has reported that more than half of their total GHG emission was due to business travel. Of these, 94.3\% are caused by flights, only 4.6\% by car journeys and 1.1\% by rail travel \cite{ethzurich}. Similarly EPFL estimated that one third of their \co2\ emissions was due to air business travels, with  87\% due to travels done by plane \cite{epfl}. Interestingly enough, a study of travelers at the University of British Columbia revealed that Academic air travel had a limited influence on professional success \cite{WYNES2019959}.

Scientists and in particular astronomers and astrophysicists, who are often exposed to the public and who can also talk about the rather unique place of the Earth in the Universe and its long evolutionary sequence, must lead by example and take actions to reduce their environmental impact. This issue is now actively discussed within the science community and ways of doing astronomy in a low Carbon future are now being proposed for immediate implementation \cite{williamson2019arxiv,matzner2019arxiv,stevens2019arxiv}. The mobilisation goes obviously well beyond the astronomical community, and there are a large number of initiatives in academia, questioning the way to continue research, while minimizing its environmental impact, e.g \cite{epfl,ethzurich,lequere2015,Flying2019,flyinglesinacademia,Labos1point5} to list a few resources and initiatives. 

In this context, it is worth looking at the travel footprint associated with the development of a large astronomy project, such as the X-ray Integral Field Unit (X-IFU): the cryogenic spectrometer of the flagship Athena Space X-ray observatory of the European Space Agency  \cite{barret2018spie}. The X-IFU consortium currently involves 13 countries, 11 in Europe plus Japan and the United States. It can thus be anticipated that the X-IFU travel footprint is large, and actions to reduce the project footprint must be put in place and have their impact quantified. Those actions must be well thought as not to impact the development of the project. Face-to-face meetings are often necessary to solve technical issues, to build the instrument in sequence, but also because the so-called \emph{social engineering} enables people to know each other better and feel part of a larger team which is critical for the success of long term endeavor that often represents space projects. 

In order to raise awareness on the need to change our attitude towards flying, but also for computing and monitoring the travel footprint associated with the development of the X-IFU, a dedicated X-IFU travel footprint calculator has been developed. Along the development of the calculator, it was realized that there is not a single commonly accepted method to estimate flight related emissions, with estimates that may differ by up to a factor of 5 or more, from one method to another. Hence, the X-IFU calculator provides estimates from seven different methods, including some of the most widely used, from either national governmental agencies or Carbon offsetting companies. In interacting with members of the community, the need for a transparent calculator became obvious, and the calculator was automized and developed as an easy-to-use web application.

The paper is organized as follows. In section \S\ref{methods}, I describe the main features of the X-IFU travel footprint calculator. In section \S\ref{otherapplications}, I describe two illustrative applications of the calculator: one for the Fall meeting of the AGU and another one for the four lead author meetings of the WGI of the IPCC. In sections \S\ref{applicationtoxifu} and \S\ref{integrated_travel_footprint}, from a representative set of meetings and travels, I compute the overall travel footprint related to the X-IFU project. In section \S\ref{wayforward}, I discuss the ways to reduce the travel footprint of the X-IFU. The conclusions are listed in \S\ref{conclusions}. Appendices A in \S\ref{appendixa} and B in \S\ref{appendixb} provide more information on the calculator methodology and the data and methods used in this paper.

\section{The X-IFU travel footprint calculator} \label{methods}
\subsection{Overview}
The tool computes the travel footprint associated with round trip flights, according to the data and methodology of several publicly available calculators. It does so for a set of trips from a given city of origin to a set of destinations. If multiple destinations are provided, the tool ranks the destinations according to the associated carbon footprint. Similarly, the tool allows us to compute the travel footprint of a large set of trips, e.g., corresponding to a conference, a meeting\ldots. For this, the city of departure for each participant to the event has first to be provided. The tool ranks the cities of departure according to their associated footprint (see Fig. \ref{dbarret_f11} for the graphical representation). If multiple destination cities are provided, the tool ranks the cities of destination according to the associated travel footprint, as to identify the host city associated with the minimum footprint, hereafter referred to as the site of minimum emission.

While most online calculators enable to compute the footprint of a limited number of trips, this tool enables us to compute the footprint of a larger number of trips in an automated way. It is therefore suited to estimate the travel footprint associated with the development of large projects, involving a lot of traveling. 

The originality of the X-IFU calculator lies in that it provides an estimate based on seven different methods, whose results are found to differ significantly. If more than one method is selected by the user, the tool returns the mean of the estimates of all selected methods.  A travel footprint calculator with similar functionalities as the X-IFU one was developed by \cite{lhackel}, considering only the \defra\ emission coefficients (see below). In addition, the X-IFU calculator enables us to compute the travel footprint associated with train journeys, assuming a predefined set of minimum distances for flying, ranging from 100 km to 1000 km. This enables us to evaluate the benefits of traveling by train in a direct way.
\subsection{Which methods are used?}
The tool incorporates data from seven different sources. Those have been selected somewhat arbitrarily, as being some of the most frequently used, referred to or commented on. For some of them, the methodology used is rather well documented, e.g. \atmosfair, \defra, \icao, \myclimate. An introduction to the general principles of the methodology used by those calculators is presented in e.g., \cite{jardine}. In alphabetic order, the data considered are from: 
\begin{itemize}
	\item \ademe: the French Environment \& Energy Management Agency with the mean emission coefficients taken from the so-called Carbon database \cite{ademe}
	\item The French Ministry of Ecology and Inclusive Transition \cite{ministryecology}
	\item \atmosfair: a German Carbon offsetting non-profit organisation \cite{atmosfair}
	\item \defra: UK Department for Environment, Food \& Rural Affairs: the UK government department responsible for safeguarding the natural environment in the UK \cite{defra}. As an example, the \climatecare\ carbon offsetting company uses the \defra\ emission coefficients.
	\item \icao\ \cite{icao}. This is a widely used on-line calculator. China Airlines for instance adopts the \icao\ methodology.
	\item KLM Carbon compensation service data \cite{klm}
	\item \myclimate: a Carbon offsetting non-profit organisation, used in particular by Lufthansa \cite{myclimate}
\end{itemize}

As stated above, the list is obviously not exhaustive but represents a variety of estimates from low to high values. Yet the web application has been designed in such a way that it enables new methods to be added. \\

\begin{figure}
\centerline{	\includegraphics[width=0.8\linewidth]{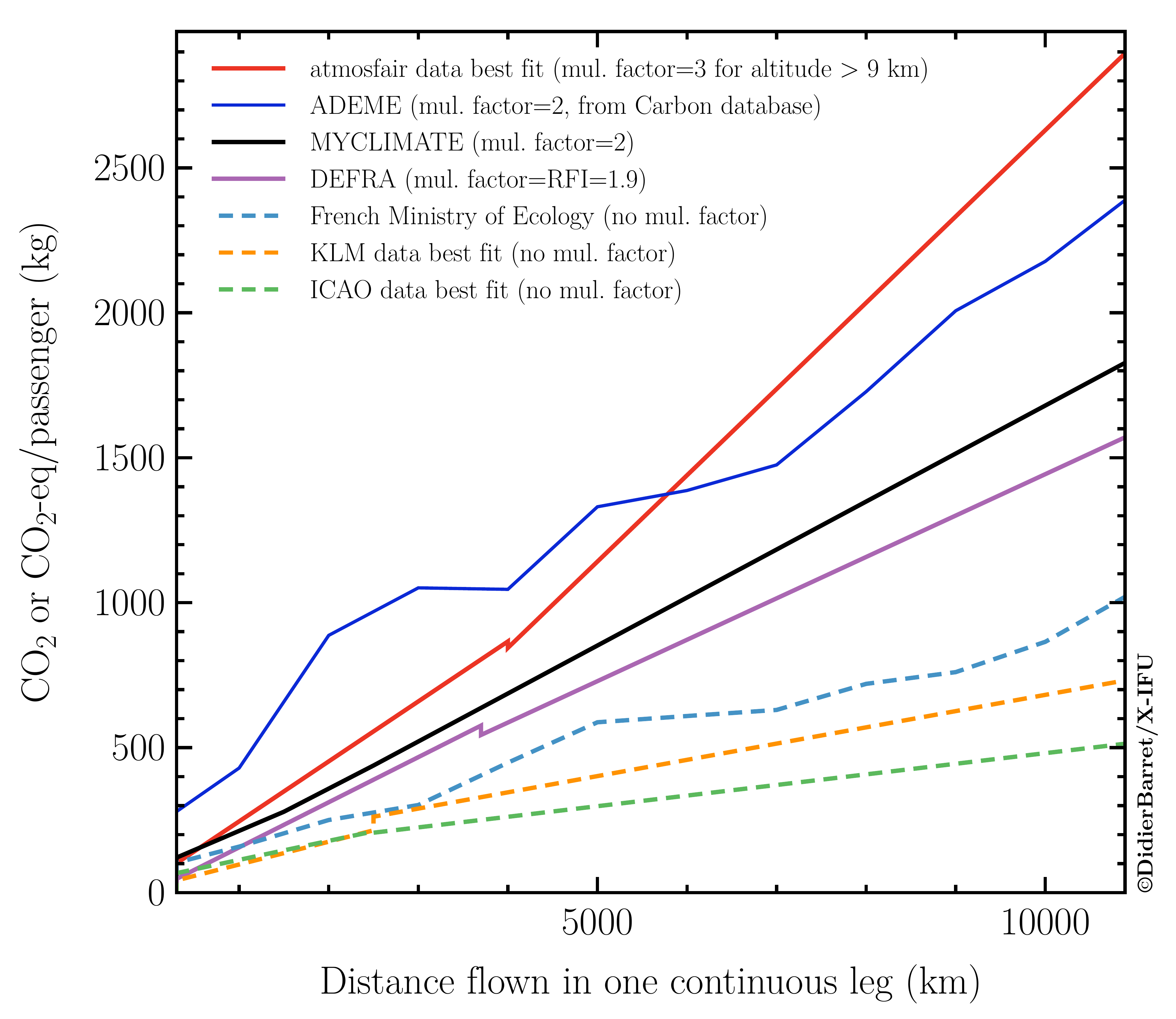} }
	\caption{Emission factors (\co2\ in dashed lines, or \co2eq\ in continuous lines, per passenger in kg) as a function of the distance flown in one continuous leg (km) as derived from the seven data sources used by the X-IFU calculator. The emission is computed assuming economy seating. Jumps in the functions are related to boundaries of interval distances over which the fit is applied or the mean emission factors defined. } 
	\label{dbarret_f1} 
\end{figure}

ADEME, \defra\ and the French Ministry of Ecology provide mean emission factors, as a function of flight distance. \myclimate\ provides an analytical formula. For \icao\ and \atmosfair\, the on-line calculators have been run for a wide range of flights of varying distances ($\sim 100$ flights spanning from 300 km to 12000 km) and the estimates have been fitted with linear functions, covering adjacent distance intervals. For its Carbon compensation service, KLM provides on its website a table of emissions for a large range of KLM flights. The KLM data have also been fitted with linear functions. Deviations by up to a few tens of percent may be found between the data and the linearly interpolated values, but those deviations are much smaller than the differences in the estimates of the various methods considered (see Appendix B in \S\ref{appendixb} for further details on the different methods). The emission functions for the seven methods are shown in Figure \ref{dbarret_f1}. 

\subsection{Accounting for non-\co2\ effects}
Emission factors or travel footprint calculators account for or ignore non-\co2\ effects. The \icao\ calculator does not account for any non-\co2\ effects awaiting for the scientific community to settle on the issue\footnote{\url{https://www.icao.int/environmental-protection/CarbonOffset/Pages/FAQCarbonCalculator.aspx}}. The same appears to be the case for the emission factors of the French Ministry of Ecology and Inclusive Transition and the KLM data, as the estimates provided are close, although a little higher than the ones of \icao\ (see \S\ref{difference_in_emission_factors} for a discussion on parameters entering the emission factors). The methods which do not account for non-\co2\ effects are plotted with dashed lines in Figure \ref{dbarret_f1}. \defra\ provides emission factors with and without a multiplier. Here we use the \defra\ emission factors, which include a multiplication factor, referred to as a radiative forcing index, of 1.9 to account for non-\co2\ effects, noting however that \defra\ raises a warning on the uncertainty associated with the latter value \cite{defra}. A 2.0 multiplier is considered by \myclimate, \ademe, and \atmosfair\ considers a multiplier of 3 for all emissions above 9 km, accounting for the profile of the flight (for long distance flights, this means a multiplier of $\sim 2.8$, see Figure \ref{dbarret_f13}). The methods applying a multiplier are shown with continuous lines in Figure \ref{dbarret_f1}.

Accounting for a multiplier, the multiplied direct flight emission is expressed in units of kg of \co2eq.

\subsection{Other parameters entering the emission factors}\label{difference_in_emission_factors}
As shown in Figure \ref{dbarret_f1}, differences by up to a factor of $\sim 5$ are found between the different data providers for flights of 10000 km legs. It is beyond the scope of this paper to discuss each method in details, but it is clear that, besides the multiplier assumed by the data providers, there are other reasons why the emission factors differ from one method to another. All estimates are computed assuming economy seating, so that cannot explain the observed differences. Parameters entering the emission factors are:
\begin{itemize}
\item the actual fuel consumption per aircraft kilometer (which depends on the plane and engine type, age, the way the plane is operated in flight, the take off and landing sequences, including taxying\ldots), 
\item the fleet considered, 
\item the correction for deviations from the great circle distance (due to holding patterns, avoidance of bad weather conditions\ldots), 
\item the assumed seating configuration of the plane and the weighting by seating class, 
\item the passenger load factor,
\item the fraction of the fuel burnt allocated to the freight, 
\item the addition or not of the emission related to the pre-production (refinery) and transport of the fuel used by the aircraft\footnote{\myclimate\ assumes that the emission factor for combustion of jet fuel (kerosene) to 3.15 kg \co2eq/kg jet fuel and the factor for pre- production used here is 0.538 kg \co2eq/kg jet fuel.}, 
\item the addition or not of the emission related to the fabrication, maintenance and disposal of the aircraft, 
\item and finally the emission related to the airport infrastructure itself. 
\end{itemize}

Let us take the illustrative example of the seating configuration as a key parameter entering the emission factors. On an Airbus A380, the seating capacity ranges from 525 passengers split between the first, business and economy classes to more than 800 in a single-class economy class layout. Depending on the way the plane is assumed to be filled the \co2\ emission factor per passenger.km could differ by about a factor of 1.5 (that may explain in part why the \icao\ estimates are lower than the others in Figure \ref{dbarret_f1}, because it makes the assumption that all aircrafts are entirely configured with economic seats). The parameters entering the emission factors are not always available for the methods used by the calculator, while one would naively expect that they are somewhat standardized and known to some accuracy. However, this is clearly not the case and the way they are considered by the data providers must explain the remaining factor of 2-2.5 difference, on top of the different assumptions they consider for accounting for non-\co2\ effects. In addition to the need to settle on the issue of the accounting for non-\co2\ effects, this clearly calls for a neutral organization to define a commonly accepted methodology for computing aircraft emissions, even more so because the overall demand for flying is predicted to increase significantly over the next decades, with little hope to make it more energy efficient. 

\subsection{Accounting for train emission}
The minimum distance considered for flying (one leg of the round trip computed as the great circle distance between the origin and destination cities) is an input to be selected by the user (it is set to 500 km by default). Below the minimum distance for flying, it is assumed that the journey is done by train. The tool then computes the travel footprint associated with train journeys. The train travel distance between city pairs is computed from the great circle distance multiplied by a factor of 1.35. This factor was derived from the comparison between road distances, assumed to be a proxy of the train travel distances, and great circle distances (see section \S\ref{road_distance_vs_gcd}). 

For the train emission factors (covering operations only), the calculator assumes the mean of the emission factors of national and international rails, as provided by \defra\ (i.e. 23.1 grams of \co2eq\ per passenger.km) \cite{defra} (more about the assumptions used by \defra can be found in \S\ref{defra_train_emission_factors}). It should be stressed however that factors for the train emission will depend strongly on the way the train is powered.  For instance, the French emission factors provided by \ademe\ are 3.4 and 5.1 grams of \co2eq\ per passenger.km for high speed trains and normal trains respectively. This rather low value is explained by the fact that electricity production is mainly associated with low carbon emission (nuclear, hydropower, renewable, with only a small fossil fuel fraction). On the other hand for a train powered by Diesel, the emission may approach $\sim 80$ grams of \co2eq\ per passenger km\footnote{A train footprint calculator for four different ways of powering trains from low \co2\ electricity (hydroelectric, sun, wind\ldots) to Diesel is available at \url{https://www.engineeringtoolbox.com/CO2-emissions-transport-car-plane-train-bus-d_2000.html}. The calculator assumes a mean emission of $\sim 22$ grams of \co2eq  per passenger.km when the energy source is an even blend of non-emission electric power and power from power plants fired with hydrocarbons (gas, oil or coal).}. 

Adopting the mean \defra\ value for train and the \defra\ emission factor for short-haul flights, a round trip Amsterdam-Munich ($\sim 1300$ km in total) will generate $\sim 40$ kg and $\sim 220$ kg of \co2eq\ by train and plane respectively.

\section{First applications of the X-IFU travel footprint calculator}
We now consider two applications of the travel footprint calculator to illustrate its capabilities.
\label{otherapplications}
\subsection{The case for a large international conference: the annual meeting of the American Geoscience Union (AGU)}
The annual meeting of the AGU is one of the largest scientific conferences, gathering in San Francisco more than 24000 scientists coming from over one hundred countries. The footprint of the 2019 fall meeting was carefully estimated to be $\sim 69000$ tons of \co2eq\ by \cite{klover}, who also evaluated the benefits of virtual participation from the most distant attendees. It was found that emissions could  be reduced by 76\% provided that the 36\% highest emitting attendees (from almost every country outside of North America) would participate virtually. Similarly, holding the conference in 3 hubs (Chicago, Seoul, Paris), and no virtual participation would result in a saving of $\sim 70$\% \cite{klover}.

As a sanity check, we run the X-IFU footprint calculator from the list of cities of departure provided by \cite{klover}. Considering the \atmosfair\ method, consistent with the high emission factors used by \cite{klover}, and assuming a minimum distance for flying of 300 km, we estimated the total emission of the meeting to be $\sim 68300$ tons of \co2eq, i.e. a value consistent with the one of \cite{klover}. Considering the \icao\ method, the total emission would reduce to $\sim 14800$ tons of \co2. Running the X-IFU footprint calculator with the average of the \ademe, \myclimate, and \defra\ estimates, and assuming train journeys for great circle distances less than 700 km, the total emission from the meeting, amounts to $\sim 50500$ tons of \co2eq. This clearly shows how sensitive the estimate travel footprint is to the assumptions used by the calculator, with a factor of 5 difference between the low and high estimates. 

\begin{figure}[!h]
	\centerline{\includegraphics[width=0.95\linewidth]{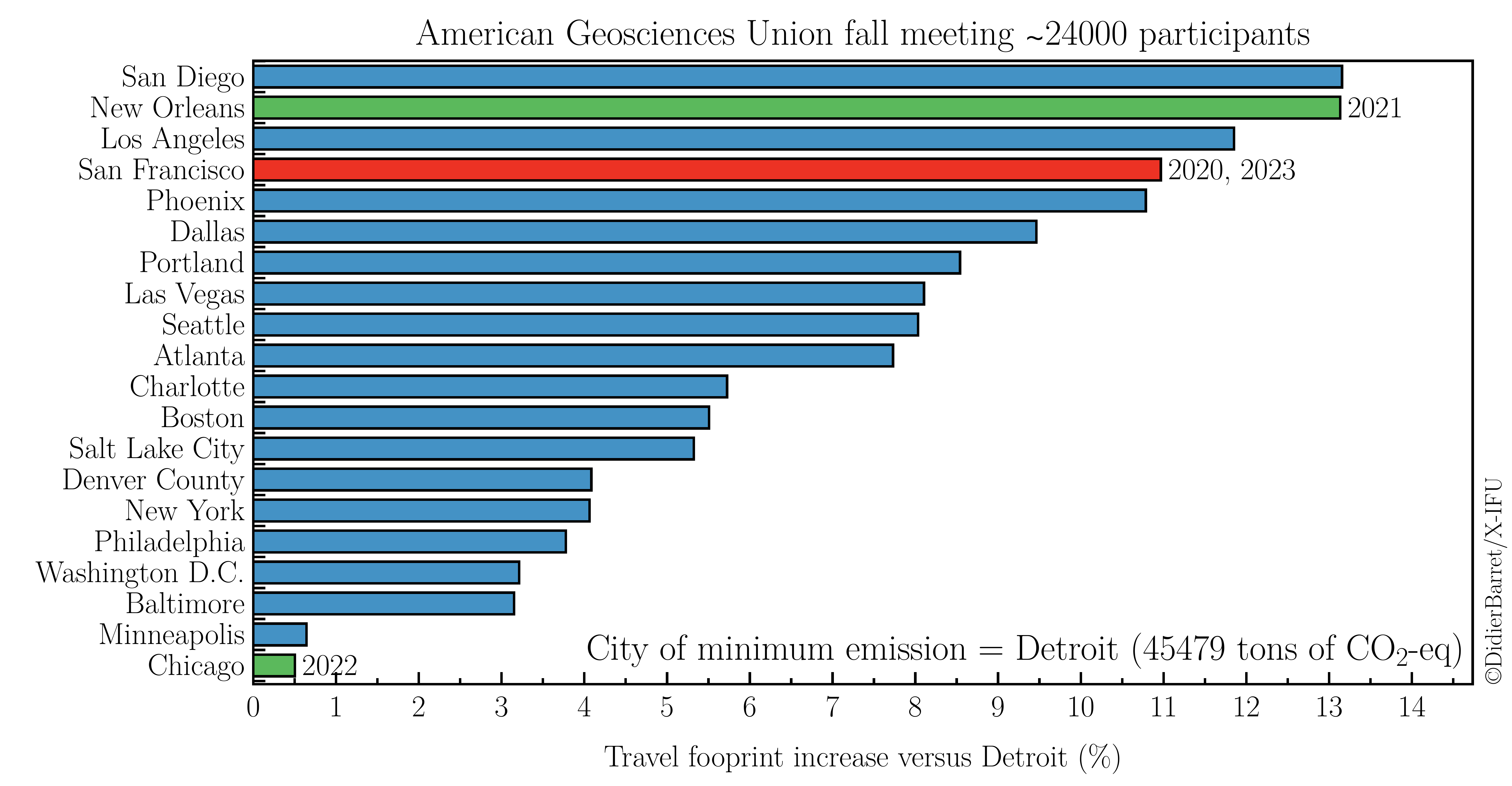}} 
	\caption{Percentage increase of travel footprint with respect to the city corresponding to the minimum footprint (Detroit), for cities hosting a large international airport in the United States. The X-IFU footprint calculator was run with the average of the \ademe, \myclimate, and \defra\ estimates, assuming train journeys for great circle distances less than 700 km. With these assumptions, for the AGU 2019 Fall meeting, the city minimizing the overall footprint would have been Detroit (while the meeting was held in San Francisco). The total footprint associated with San Francisco was $\sim 5000$ tons of \co2eq higher (11\%). The four upcoming meetings are San Francisco (2020), New Orleans (2021), Chicago (2022) and San Francisco again in 2023. They are highlighted by green and red bars. Assuming the same audience as the 2019 Fall meeting, the travel footprint associated with New Orleans for the 2021 meeting would be 13\% higher than a similar meeting held in Detroit.} 
	\label{dbarret_f2} 
\end{figure}

We have also run the calculator to search for the site of minimum emission in the United States, requiring potential city hosts to have at least one large international airport. With the above assumptions, the city of minimum emission would be Detroit, with a total footprint of $\sim 45500$ tons of \co2eq. The footprint associated with Chicago and Minneapolis are within $\sim 0.5$\% of Detroit (and Chicago would be the city of minimum emission for an audience restricted to the $\sim 14$ thousands US participants). This means that the travel footprint associated with San Francisco is $\sim 11$\% higher than the one of Detroit (see Figure \ref{dbarret_f2}). The next four AGU December meetings will be held in San Francisco (2020), New Orleans (2021), Chicago (2022) and San Francisco again in 2023. If attended by an audience similar to the 2019 meeting, one can already anticipate that New Orleans would increase the footprint by $\sim 13$\%, compared to Detroit or Chicago, and $\sim 2$\%, more than San Francisco (about 1000 tons of \co2eq\ more, see Figure \ref{dbarret_f2}).  

Let us now get some feeling about what these large numbers actually mean. As a global progress report on climate action, the 2019 Emissions Gap report \cite{emgr2019} issued by the United Nations Environment programme stated that in 2018,  the total greenhouse gas emissions reached a peak of 51.8 Gt\co2eq (excluding emissions from land-use change), with emission growing steadily at a rate of 1.5\% over the last decade. For a world population of 7.5 billions, this translates to an average emission of $\sim 7$ tons of \co2eq\ per capita. With this number, the total footprint of the 2019 Fall meeting of the AGU ($\sim 50500$ tons of \co2eq) is equivalent to the annual footprint of about 7200 world citizens, but released within a week. Alternatively, on average, each of the 24000 participants increased her/his own footprint by $\sim 2.1$ tons of \co2eq\ by attending the meeting.

\subsection{IPCC Working Group I AR6 Lead Author meetings}
 The IPCC Working Group I (WGI) aims at assessing the physical scientific basis of the climate system and climate change. It is currently in its Sixth Assessment cycle, leading to the sixth Assessment Report (AR6). This assessment is done by authors from various regions of the world, and supported by bureau members and a Technical Support Unit (partly based in Paris Saclay, France, and Beijing, China, where co-chairs are located). As part of the preparation of the AR6, four Lead Author Meetings (LAM) are needed: one took place in Guangzhou (China, 215 participants), another one in Vancouver (Canada, 222 participants), another one in Toulouse (France, 248 participants) and the last one is planned in Santiago  (Chile, 304 participants). In an inter-governmental organization such as the IPCC, there is a systematic search for balance between regions of the world where meetings are hosted. The location of Lead Author meetings results from proposals from governments, cities, and universities or research centers. Here we are interested to compare the travel footprint of the four meeting locations, with respect to their attendance. The participant list to each of the four LAM was provided by the Working Group I Technical Support Unit. For illustrative purposes, the travel footprint calculator was first run, considering the \icao, \defra, and \atmosfair\ estimates, i.e. from the lowest to the highest estimates, and considering a minimum distance for flying of 700 km, equivalent to accepting train journeys up to $\sim 10$ hours or so. Summing the footprint of the 4 meetings, the estimates of \icao, \defra, and \atmosfair\ are $\sim 870$ tons of \co2\ and $\sim 2700, 4940$ tons of \co2eq\ respectively, meaning a factor of $\sim 5.5$ difference in the computed emissions between the low (\icao) and high (\atmosfair) estimates. This range simply reflects the different emission functions reported in Figure \ref{dbarret_f1}.

\begin{figure}[!h]
	\centerline{\includegraphics[width=0.495\linewidth]{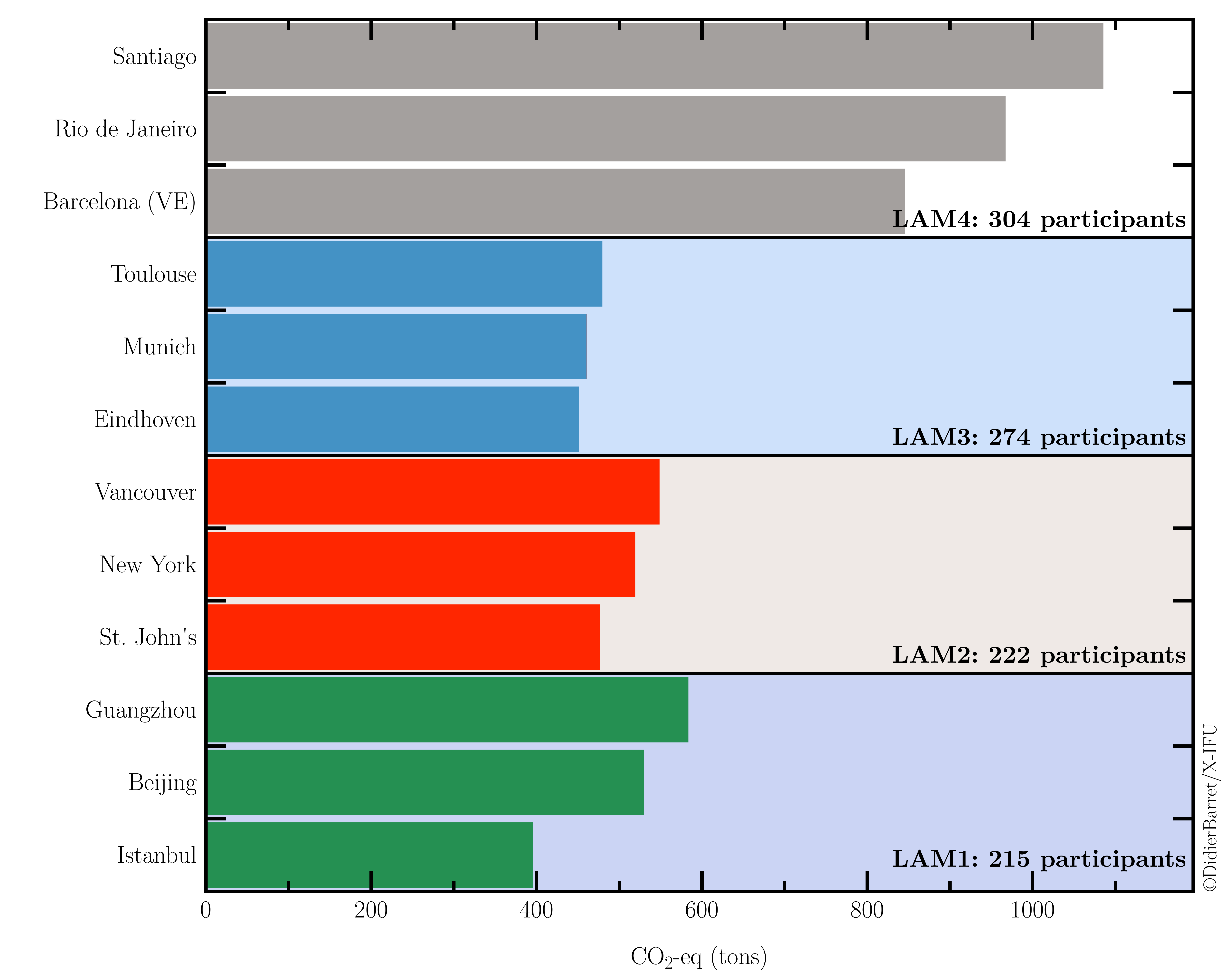}\includegraphics[width=0.495\linewidth]{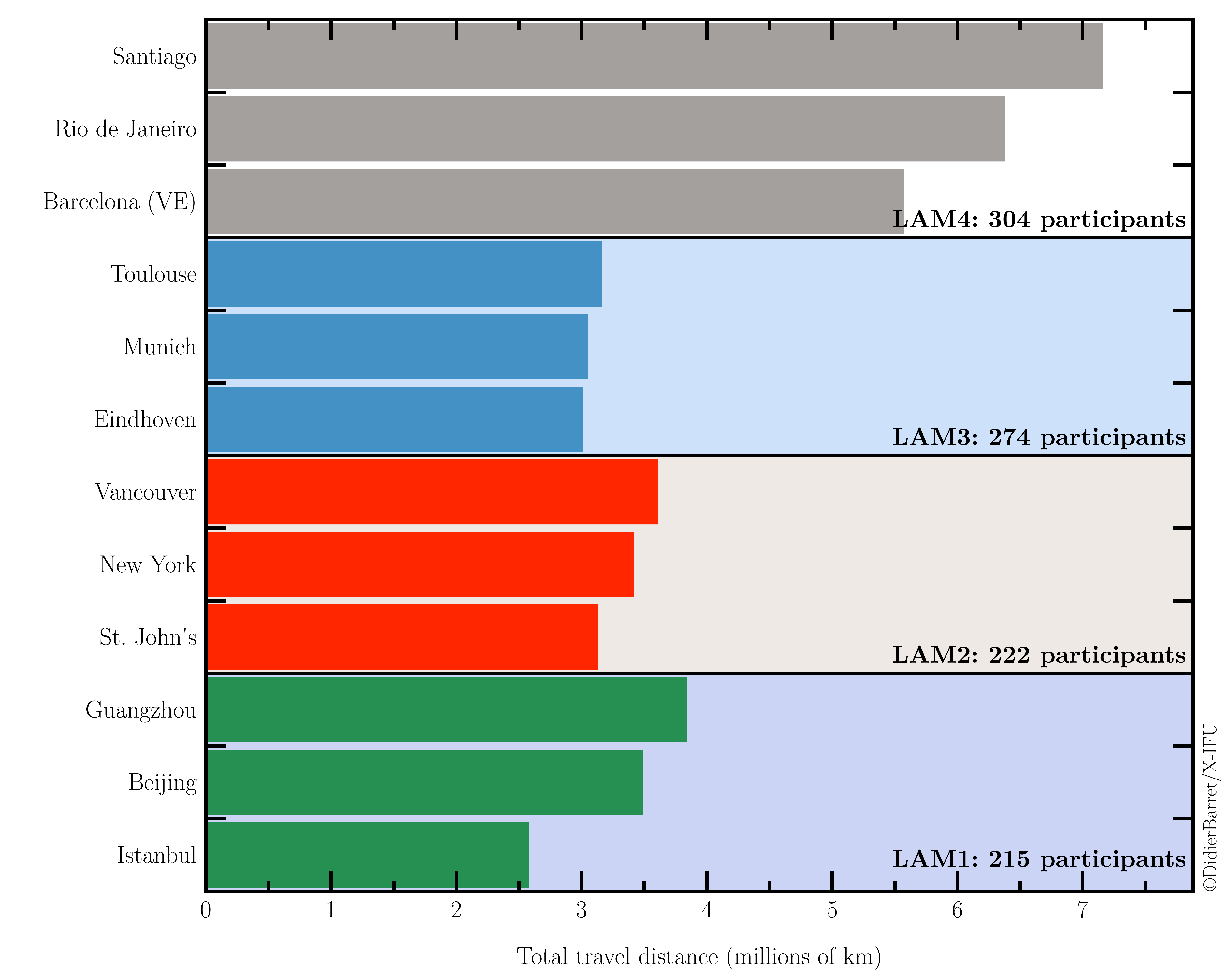}} 
	\caption{Left) The travel footprint associated with the 4 IPCC WGI Lead Author Meetings, based on the \defra\ emission factors and assuming a minimum distance for flying of 700 km. The footprint is computed at the city of minimum emission (bottom bar of each sub-panel), for an intermediate city host (middle bar of each sub-panel), and finally for the host city of the meeting (top bar of each sub-panel). Right) The total travel distance to reach the various city hosts is plotted in millions of km. Barcelona refers to the city of Venezuela.} 
	\label{dbarret_f3} 
\end{figure}

The four LAM meetings ought to be located in four different continents, namely Asia, North America, Europe and South America. The calculator was run to identify the city of each continent that would be associated with the minimum emission given the audience of each meeting. For this purpose, a list of cities hosting large international airports in each continent, was given as an input to the calculator as the list of potential city hosts of the LAM. The calculator was run assuming the \defra\ emission factors and considering a minimum distance for flying of 700 km. The results are shown in Figure \ref{dbarret_f3}. Because for all four LAM, there is a relatively large number of participants from Western Europe, reflecting a geographical imbalance in the authors of the AR6 WGI Report linked to an existing imbalance in ongoing world climate science research activities, the best meeting locations is expected to be on the western part of Asia and the eastern part of South and North America. This is obviously what shows Figure \ref{dbarret_f3}. For example, for the meeting in Asia, Istanbul would have reduced the emission from 584 tons to 400 tons of \co2eq ($\sim 30$\% reduction). Constraining the meeting to be in China, Beijing would have reduced the emission to 530 tons of \co2eq ($\sim 10$\% reduction). Similarly, for the LAM4, having it located at Barcelona (Venezuela) would have saved $\sim 20$\% of the emission compared to Santiago. For the other two meetings (LAM2 and 3), the city selected was within $\sim 10$\% of the city of minimum emission. For indication, the total travel distance of the 4 LAM is also shown in Figure \ref{dbarret_f3}. It varies from $\sim 3$ (Toulouse) to $\sim 7$ (Santiago) millions of km. 

Translating again the above numbers to meaningful values, if one assuming a total footprint of the four LAM of 2700 tons of \co2eq (i.e. the intermediate \defra\ estimate), spread over four weeks, this is equivalent to the annual footprint of about 400 people (considering a mean annual emission per capita of $\sim 7$ tons of \co2eq\, \cite{emgr2019}). Alternatively, for each meeting, on average, each participant increased its own footprint by $\sim 2.7$ tons of \co2eq.

The above example shows the power of having such a travel footprint calculator for optimizing the meeting locations of recurring events of similar attendance. The calculator can obviously be run for optimizing the location of events worldwide, across continents, geographical regions\ldots. Obviously in large international inter-governmental organizations, such as the IPCC, political, economical, ethical, visibility arguments (including the need to support developing countries) must necessarily enter  into consideration when selecting meeting locations. The calculator presented here may still help to guide the optimum choice within such boundary conditions. 

\section{Application to X-IFU related travels}\label{applicationtoxifu}

As stated earlier, the X-IFU travel footprint calculator was first developed to provide X-IFU consortium members with a tool to easily access the footprint associated with their X-IFU related travels. It ought to be used for those organizing meetings related to X-IFU to find the optimum location, being the site of minimum emission. It was developed under the assumption that facing real numbers would be more efficient to convince people to change their habits than just talking about the mandatory need to do so. The tool is now provided as part of the registration form to consortium meetings, where each member can also easily see the benefit of traveling by train. It is now used to monitor the overall footprint of all X-IFU related travels. 

\subsection{X-IFU Consortium meeting with 120 attendees}
\begin{figure}
	\centerline{\includegraphics[width=0.8\linewidth]{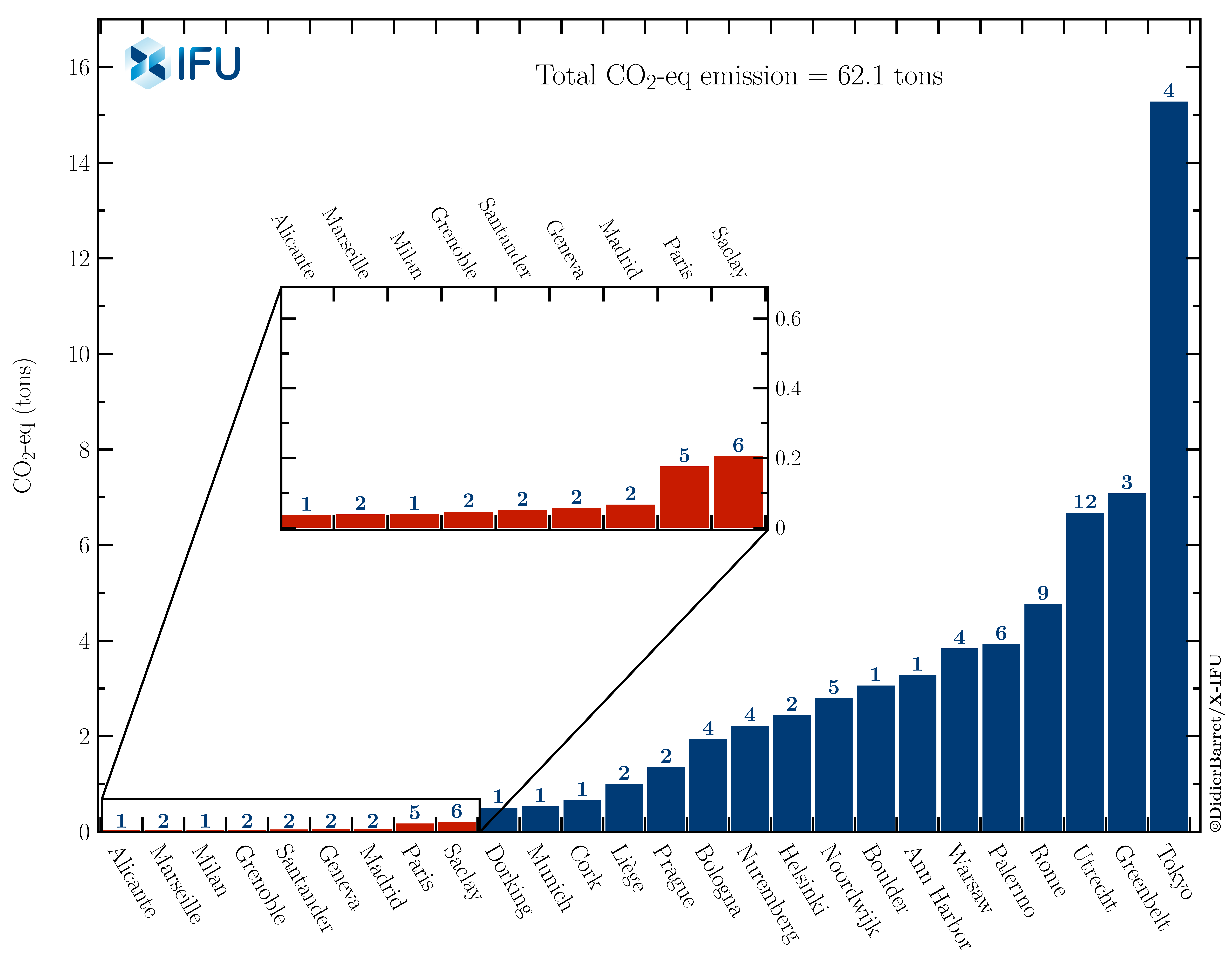}}
	\centerline{\includegraphics[width=0.8\linewidth]{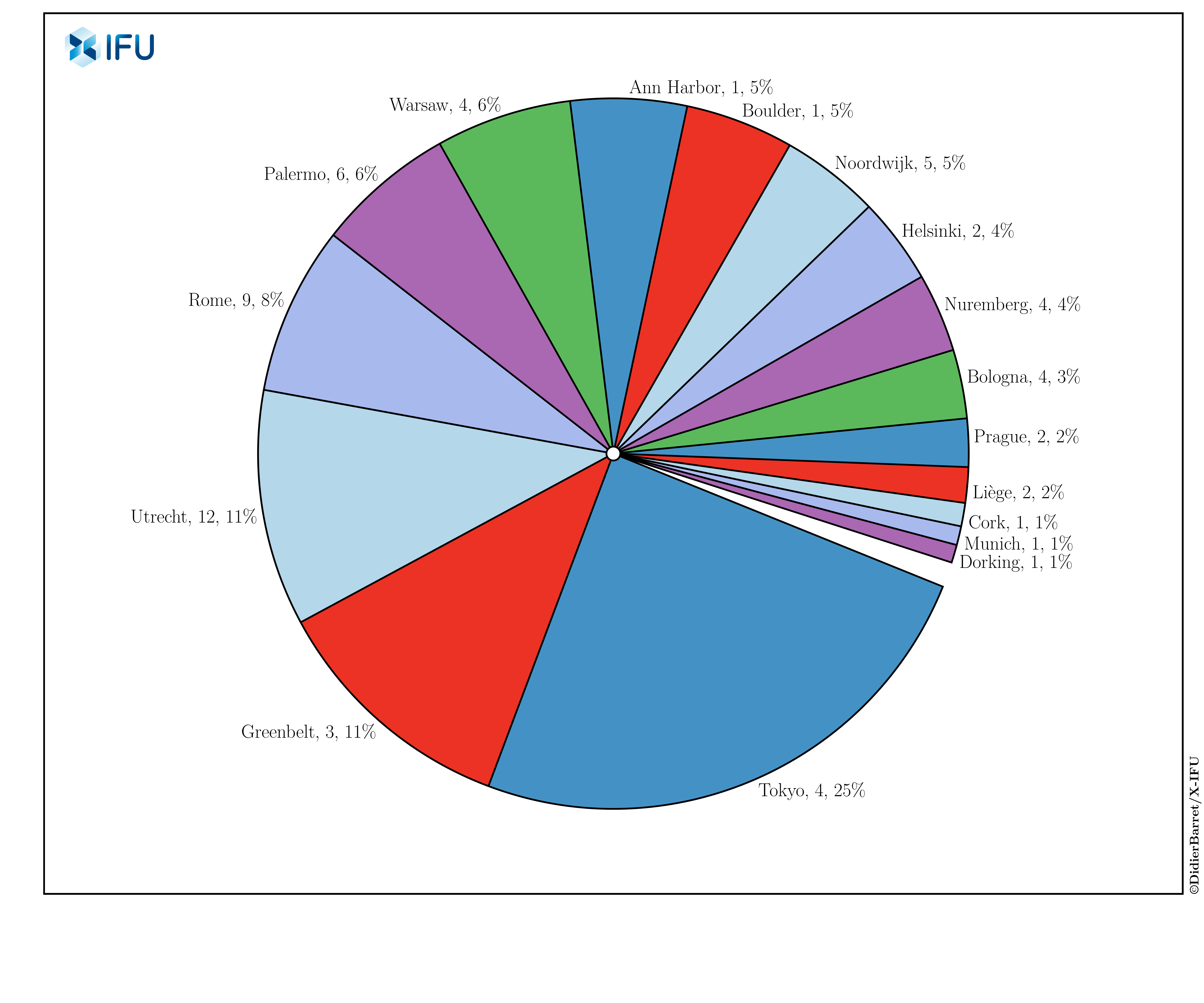}} 
	\caption{Top) The travel footprint of an X-IFU consortium meeting held in Toulouse. The carbon dioxide equivalent emission is provided for each city of origin. Identical trips (i.e. identical origins) are summed. The number of trips per city of origin is indicated at the top of each bar. The minimum distance for flying is 700 km. The travel emission associated with train journeys is indicated by red bars (see also the inset). The footprint is computed as the mean of the \ademe, \myclimate, and \defra\  estimates. Bottom) The split of the total travel emission as a function of the city of origin for the travelers. The number of attendees per city of origin is shown after the name of the city, before the percentage of the emission.} 
	\label{dbarret_f4} 
\end{figure}

Let us now first compute the travel footprint of a Consortium meeting, similar to the last one that was held in Toulouse. It was attended by 120 people originating from 26 cities, apart from Toulouse. The next one is planned to be in Li\`ege (April 2020), with the clear recommendation to travel by train whenever possible\footnote{After the submission of the paper, because of the covid-19, it was decided to turn the Consortium Meeting \#11 into a full virtual meeting. Ajustements were made to the program to maximize remote participation.}. Every other consortium meeting will be located in France. For the September 2019 Consortium meeting, in total 85 people traveled to Toulouse (the remaining 35 attendees are either from CNES or IRAP). Let us define the objective of a minimum distance for flying of 700 km, meaning that participants from e.g. Milan, Paris, Saclay, Grenoble, Marseille reached Toulouse by train. The X-IFU travel footprint calculator was run, assuming the mean of the \ademe, \myclimate, and \defra\ estimates, and considering such a minimum distance for flying. The footprint associated with such a consortium meeting is shown in Figure \ref{dbarret_f4}. The total emission of the meeting would be about $\sim 62$ tons of \co2eq (it would be 70 tons of \co2eq, under the assumption that no one came by train). As can be seen, 4 attendees from Japan would then represent 25\% of the travel emission, more than twice the travel emission from 12 attendees from SRON Utrecht. Considering only the \atmosfair\ and \icao\ estimates, one would get a travel footprint of $\sim 72$ tons of \co2eq\ and $\sim 22$ tons of \co2\ respectively.  

\subsection{Extended X-IFU Consortium meeting with 240 attendees - computing the site of minimum emission}
\begin{figure*}[!thbp]
	\centerline{\includegraphics[height=4.3cm]{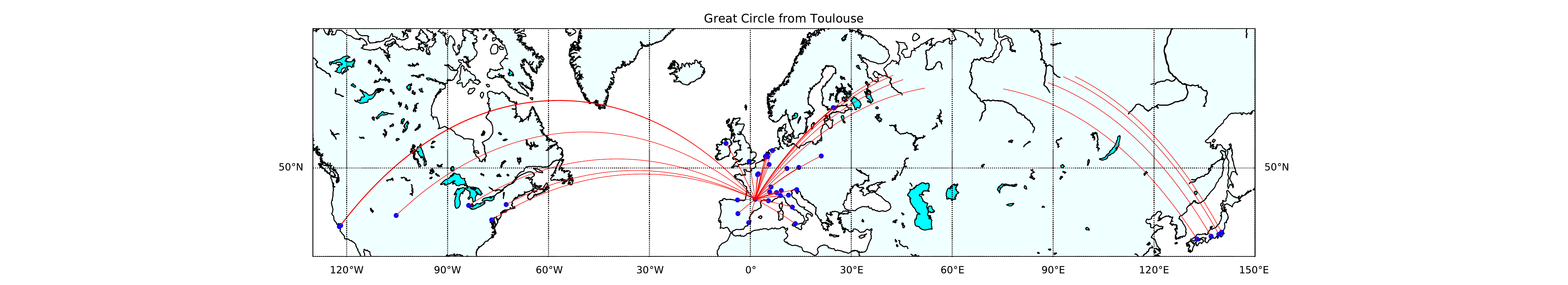}} 
	\caption{The location of the cities making up the X-IFU consortium are indicated by a blue symbol. The great circle path from Toulouse is marked with a red line. The X-IFU consortium is spread over 39 cities, in Europe, the United states and Japan. } 
	\label{dbarret_f5} 
\end{figure*}

The X-IFU Consortium includes 240 active members as of January 2020. This number is expected to grow towards the delivery of the flight model of the instrument, when there will be a significant ramping up of the hardware activities, including integration and testing. The X-IFU consortium is currently spread over 39 cities as shown in Figure \ref{dbarret_f5}.

We now wish to estimate the site of minimum emission, assuming that such extended consortium meetings are held in a city of a participating institute to the X-IFU consortium. The X-IFU travel footprint calculator was run, assuming the mean of the \ademe, \myclimate, and \defra\ estimates, and considering 700 km as a minimum distance for flying.

\begin{figure}[!thbp]
	\centerline{\includegraphics[width=0.95\linewidth]{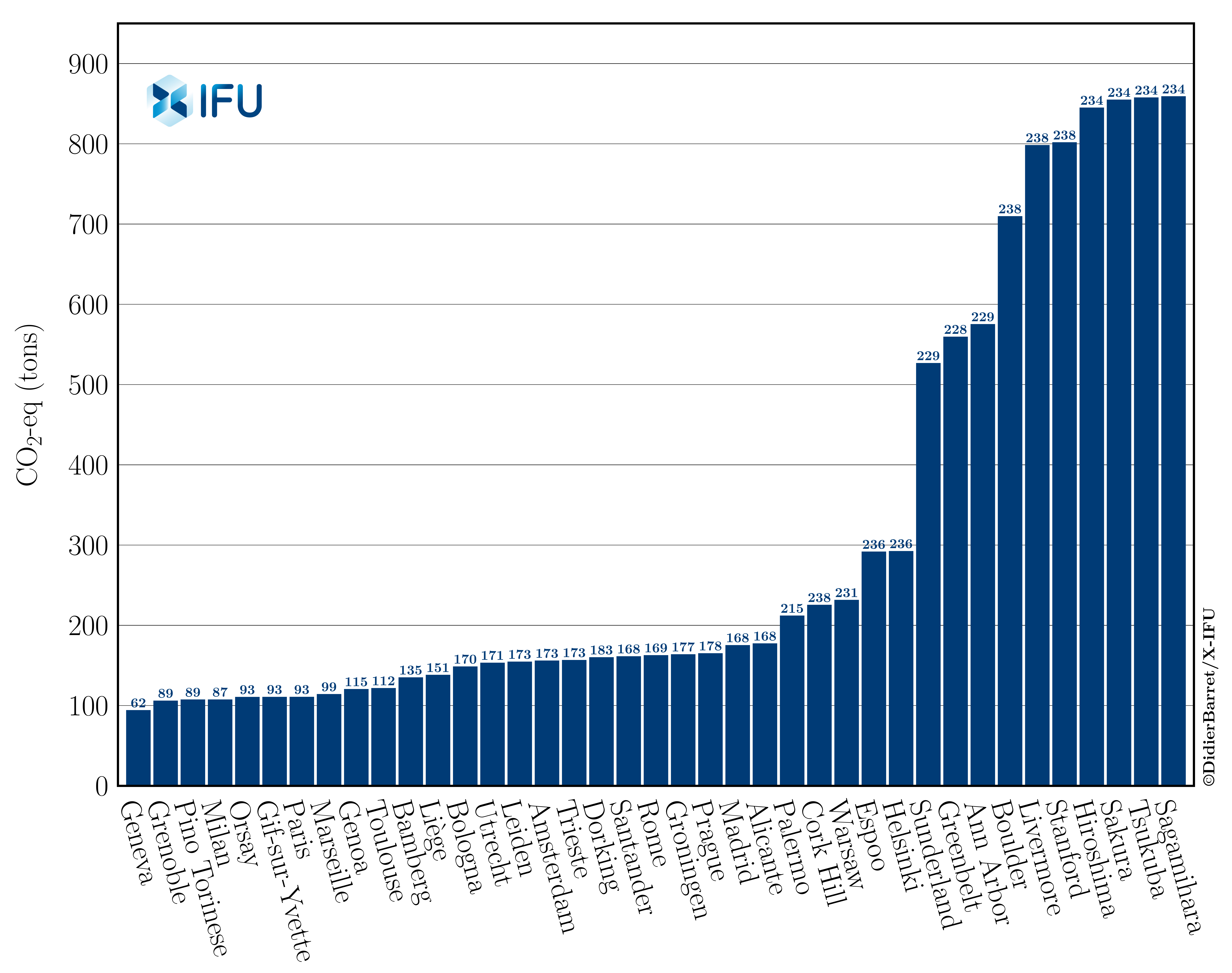}} 
	\caption{The travel footprint of consortium meetings attended by 240 participants as a function of the host city for the meeting. The X-IFU travel footprint calculator was run, assuming the mean of the \ademe, \myclimate, and \defra\ estimates, and considering 700 km as a minimum distance for flying. The number of travels by plane is indicated at the top of each bar (with the exception of the locals, all other attendees are traveling by train). The carbon dioxide equivalent emission is summed over all cities of origin and provided for each city of destination.} 
	\label{dbarret_f6} 
\end{figure}

The results are shown in Figure \ref{dbarret_f6}. As can be seen, all consortium meetings held in Western Europe, which corresponds to the geographic barycenter of the X-IFU Consortium, have a total \co2eq\ emission around 100 tons. Having a meeting on the US east coast would multiply that amount by 5-6 and having a meeting in Japan or on the West coast of the US would have an associated \co2eq\ emission 8-9 times higher, reaching about 900 tons of \co2eq\ per meeting. Such a large value would exclude the possibility to hold a meeting in the latter sites with such a large attendance. Accepting 100 tons as a maximum amount of \co2eq\ emission for an extended consortium meeting would imply that the attendance of such a meeting in Japan be reduced by a factor of $\sim 10$, assuming everyone flies in economy seats.

\begin{figure}
	\centerline{\includegraphics[width=0.8\linewidth]{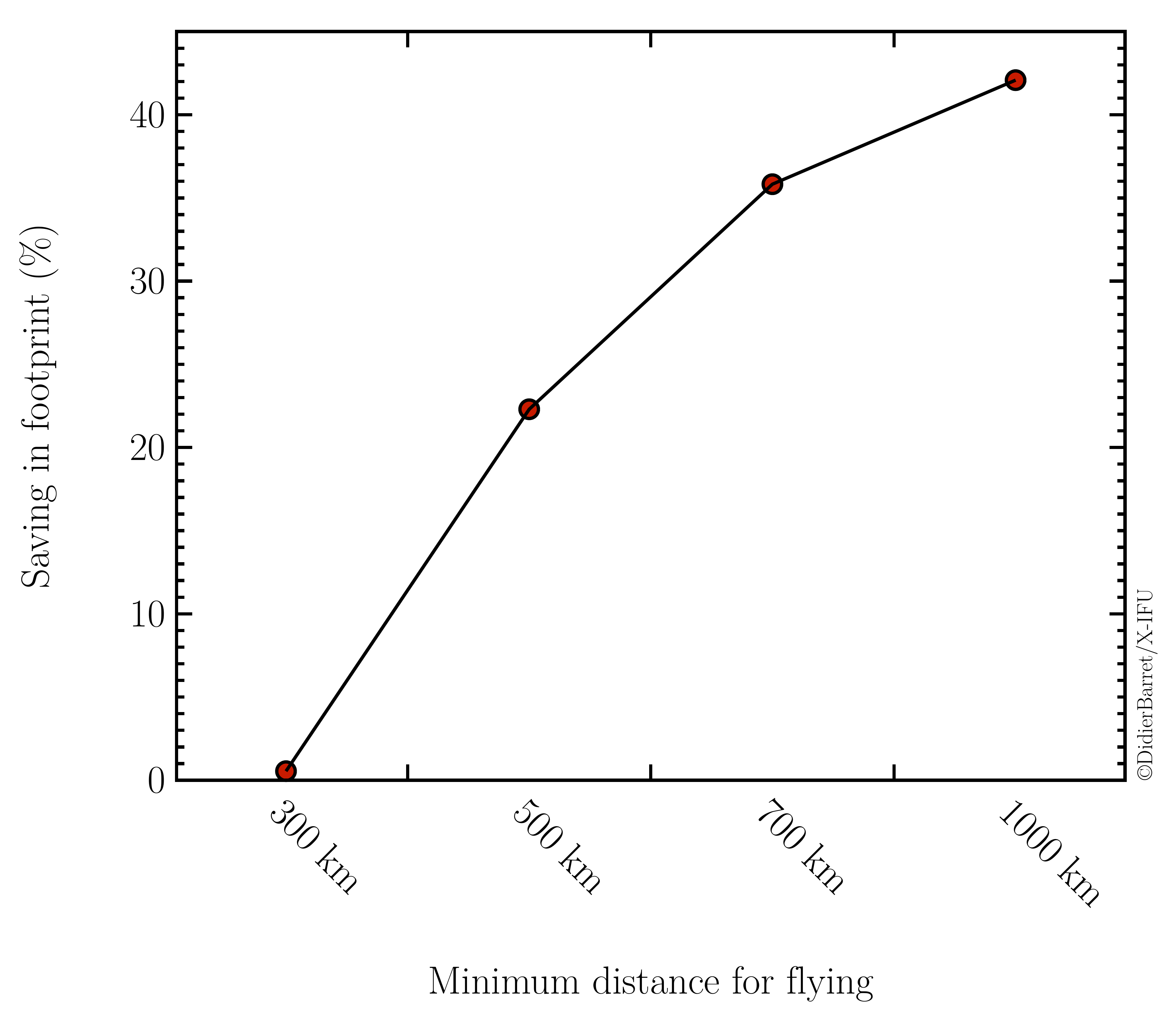}} 
	\caption{The saving in footprint as a function of the minimum distance for flying in km. The saving is against the case in which everyone flies. } 
	\label{dbarret_f7} 
\end{figure}

%
The calculator offers the option to select the minimum distance for flying from 100 km to 1000 km, which then enables us to visualize easily the savings associated with train travels. We have repeated the computation above but varying the minimum distance for flying. In the worse case scenario in which nobody travels by train, the site of minimum emission would correspond to a travel footprint of 147 tons of \co2eq. This amount would reduce by about 42\% if the minimum distance for flying is set to 1000 km. This is illustrated in Figure \ref{dbarret_f7} where we show the savings, against the case where no one travels by train for the meeting described above. Obviously, this stands for the geographical configuration of the X-IFU Consortium, but this still shows that significant savings (several tens of \%) can be achieved by considering traveling by train instead of traveling by plane. 

\subsection{Scaling to smaller X-IFU related meetings}
Assuming that the travels to consortium meetings are representative of the travels within the X-IFU consortium, from the above numbers, meaning accepting a minimum distance for flying of 700 km, one obtains a mean footprint  per travel of $\sim 120/240=0.5, 600/240=2.5, 850/240=3.5$ tons of \co2eq\ for a meeting in  western Europe, United States and in Japan, respectively. 

Two types of smaller face-to-face physical meetings are foreseen for the project. The first one is the so-called project manager meeting, held at CNES in Toulouse, and attended by up to 30 traveling participants twice a year. The project manager meetings would amount  to $2 \times 30 \times 0.5=30$ tons of \co2eq. 

The second type of face-to-face meetings foreseen are topical meetings on a specific topic (e.g. instrument sub-system interfaces). Assuming 10 such topical meetings, 8 in Europe, 1 in the US and 1 in Japan, all attended by at most 10 traveling participants, the overall emission is $8 \times 10 \times 0.5 + 1  \times 10 \times 2.5 + 1 \times 10 \times 3.5 = 35 + 25 + 35 = 95$ tons of \co2eq.  

In addition to the above meetings, one may consider that 10 key people of the X-IFU consortium, e.g. the Principal Investigator, the Project Manager, 2 Co-Principal Investigators, Chairs of the X-IFU science advisory team, Chair of the X-IFU calibration team/instrument scientist, instrument manager, performance manager \ldots will travel 10 times per year across the consortium, including once in the United States and once in Japan. This is another $10 \times (8 \times 0.5 + 2.5 + 3.5 ) = 100$ tons per year. 
\subsection{X-IFU presence to conferences}
The X-IFU and its subsystems must be presented at international conferences, e.g. to promote its scientific capabilities and keep informed the community of the progresses of the project. Within the CNES and IRAP teams (about 60 members at the end of 2019, including engineers, scientists, support staff), a coordination effort is already in place to limit the list of attendees to the bare minimum, most particularly to conferences in very distant locations (e.g. the upcoming SPIE Astronomical Telescopes and Instrumentation meeting in Yokohama, Japan).  Based on current numbers, one can estimate 30 participations to conferences in total per year for the CNES/IRAP teams (24 in Europe, 3 in Japan and 3 in United States). This translates to an annual travel footprint to conferences of $\sim 30$ tons of \co2eq per year, equivalent to an average of $\sim 0.5$ tons of \co2eq per year per member of the team. 

With the $\sim 60$ institutes and institutions involved in the X-IFU consortium and its 240 active members, it is hard to estimate precisely the associated travel footprint to conferences. It is up to each participating institute in the X-IFU consortium to define its own policy with respect to attending conferences, but there will be a strong incentive to optimize and coordinate whenever possible their presence to conferences. Extrapolating the CNES/IRAP number per personal (0.5 ton of \co2eq\ per year) to the whole consortium, one would obtain a total footprint of 120 tons \co2eq per year for X-IFU representation to conferences. Again, this number should be considered uncertain, but should also be seen as a maximum value being computed prior to any optimizations of the X-IFU presence to conferences (see discussion below). 

\subsection{Traveling for downstream activities: integration and testing}
In the current baseline, the X-IFU instrument will be integrated in Toulouse. The X-IFU today consists of about 25 independent sub-systems to be delivered by the team responsible of the procurement. There will be two instrument models before the flight model, to be integrated in Toulouse (engineering and qualification models). The staffing required for the integration sequence of these instrument models is not known yet, but it appears plausible that about 100 engineers and scientists from the Consortium will have to travel to Toulouse several times for each model. Note that each sub-system may also have its own integration sequence involving additional travels. These travels are absolutely mandatory but their footprint will also be monitored. Assuming 3 models and 100 travels per model and 3 travels to Toulouse per model integration, this would amount for an \co2eq\ emission of $3 \times 100 \times 3 \times 0.5=450$ tons (again considering that a minimum distance for flying is 700 km).  Although this number should be considered with great caution and very likely on the low side (e.g. as not accounting for trouble shooting activities, travels to support calibration\ldots), it is comparable to the one year \co2eq\ emission budget of all other X-IFU related meetings (see Table \ref{emission_budget}). 
\begin{table}
	\caption{Summary of \co2eq\ emissions for a representative set of X-IFU meetings: Consortium meetings, project manager meetings, face-to-face topical meetings, key personnel travels, participation to conferences and finally travels by consortium partners to Toulouse for integrating the X-IFU models to CNES (Toulouse). This covers the period onwards up the delivery of the flight model of the X-IFU. These numbers should be considered as maximum values, against which reduction measures are to be applied. }
	\centering
	\begin{tabular}{lcc}
	\hline
	\hline
		Meeting type & \co2eq (tons)   \\
	\hline
	Consortium meeting (120 attendees, current, one per year, up to 2023) & 60  \\
	Consortium meeting (240 attendees, max, one per year, 2024 up to 2029) & 120  \\
	Project manager meetings at CNES (30 attendees, 2 per year, up to 2029) & 30 \\
	Topical face-to-face meetings (10 people, 10 per year, up to 2029) & 95  \\
	Additional travels by 10 key consortium members (10 per year, up to 2029) & 100 \\
	Participation to international conferences (up to 2029) & 120 \\
	\hline
		Total travel emission per year & $\sim 405-465$ \\
		Travels to CNES for the integration of X-IFU instrument models & $\sim 450$ \\
	\hline
		Integrated travel footprint  (2015-2029) & $\sim 7000$\\
	\hline
	\end{tabular}
	\label{emission_budget}
\end{table}

\section{Integrated X-IFU travel footprint and what it means}
\label{integrated_travel_footprint}
The travel footprint of the X-IFU is summarized in table \ref{emission_budget}. The list of X-IFU related meetings is built upon more than five years of the existence of the consortium, hence the travel footprint budget derived, although approximative, can be considered representative. Integrated from 2015 until the year of the delivery of the flight model of the X-IFU, the total travel footprint associated with the development of the X-IFU is rounded to $\sim 7000$ tons of \co2eq. It is interesting to note that this is less than the saving of 10000 tons that would have been achieved by moving the 2019 Fall meeting of the AGU from San Francisco to Detroit/Chicago. This being said, the numbers provided in table \ref{emission_budget} are rather large, and should be seen as maximum values, against which efficient reduction measures are going to be implemented. 

So let us first consider the average annual travel footprint of X-IFU to be $\sim 500$ tons of \co2eq\, and relate this rather abstract number to quantities that can be better apprehended. According to the U.S. Environmental Protection Agency Greenhouse Gas Equivalencies Calculator\footnote{\url{https://www.epa.gov/sites/production/files/widgets/ghg-calc/calculator.html}}, 500 tons of \co2eq\ correspond to $\sim 106$ passenger vehicles driven for one year, 85 home electricity use for one year, could be avoided by recycling $\sim 170$ tons of waste or would be sequestered by $\sim 8300$ tree seedlings grown for 10 years. 

Assuming a mean worldwide emission of $\sim 7$ tons of \co2eq\ per capita \cite{emgr2019}, the annual footprint of the X-IFU project contributes equally to the greenhouse gas emission of $\sim 70$ persons. Alternatively, the X-IFU travel related emission of each consortium member accounts already for about 30\% of the current average level of emission worldwide. As stated in the Emission gap report \cite{emgr2019}, by 2030 (which is the horizon for the development of the X-IFU), GHG emissions would need to be reduced by 25\% (=4.9 tons of \co2eq\ per capita) or 55\% (=2.9 tons of \co2eq\ per capita for a projected world population of 8.5 billions) from the 2018 values to limit the global warming to below $2^\circ$C and $1.5^\circ$C, respectively. This means that in 2030, the annual travel footprint of each X-IFU Consortium member if unchanged (about 2 tons of \co2eq) would amount for $\sim 40$\% and $\sim 70$\% of a sustainable annual carbon budget.

In Europe, by 2050, in the context of necessary reductions by developed countries as a group, according to the IPCC, the European Union objective is to reduce GHG emissions by 80-95\% below 1990 levels. With current measures, the target calculations including emissions from international aviation and excluding emissions and removals from the land sector set the ultimate goal of GHG emission between $\sim 0.5$ and $\sim 2$ tons per capita at the 2050 horizon (for an assumed projected population of $\sim 523$ millions people). This gives clear markers against which the mean annual X-IFU related travel footprint of $\sim 2$ tons of \co2eq\ should be compared to, given that on average flight related travel emission is at most a few percents of total greenhouse gas emission per capita, the latter being dominated by contributions from other means of transportation, housing, food, goods and services.

\section{Reducing the X-IFU travel footprint}\label{wayforward}
Based on the numbers presented in table \ref{emission_budget}, the X-IFU Consortium board has already agreed to take actions to monitor and reduce the travel footprint of the X-IFU. The first decision taken was to reduce the number of Consortium meetings to one per year instead of two. The numbers presented in table \ref{emission_budget} are the one to work against, and to which reduction measures will be proposed for implementation within the X-IFU Consortium on the shortest possible term. Those measures have already been discussed in the context of reducing the environmental impact of scientific activities \cite{epfl,ethzurich,williamson2019arxiv,matzner2019arxiv,stevens2019arxiv,lequere2015,Flying2019,flyinglesinacademia,Labos1point5}, such as conferences, but they now have to be put in place along the development of a large infrastructure, such as the X-IFU. The actions foreseen for X-IFU are listed below:
\begin{itemize}
\item Favor video conferences whenever possible, even for meetings in which critical matters are discussed and face-to-face discussions may be preferred. As an example, analyzing the total cost of videoconferencing, including operating costs of the network and videoconferencing equipment, lifecycle assessment of equipment costs, although the footprint of virtual meetings  depend on many factors such as distance travelled, meeting duration, and the technologies used, \cite{ong} found that videoconferencing takes at most 7\% of the energy/carbon of an in-person meeting.  A video-conferencing system (e.g. Zoom) has been set-up for both the CNES and IRAP teams, and provides satisfaction, even for meetings attended by more than 40 people. It is thus likely that some topical meetings will be held by videoconference. This may have the further benefit of increasing attendance and increasing efficiency due to the time saved by not traveling. 
\item Optimize the travels of all participants to the one week long Consortium meetings by organizing topical meetings, next to the plenary sessions, as to cover a full week. On a project like X-IFU, we do need physical interactions within and outside work. This also means that special attention should be brought to the social events linked to consortium meetings.
\item Avoid short duration single goal meetings, and make the best use of each travel, extending the stay to participate to nearby events (schools, seminars,\ldots), to network, especially for the youngest members of the Consortium.
\item Keep the \co2eq\ emission of all consortium meetings around 100 tons, and restrict the attendance accordingly, but sweep the locations across the whole consortium, i.e., not always meeting in western Europe. This is required not to impose the most distant members of the X-IFU consortium to fly every time, hence being always exposed to jet lags, fatigue, losing time on travels\ldots. This is also required to ensure that each contributing country receives proper visibility.
\item Investigate the possibility to hold multi-node distributed consortium meetings: in Europe, Japan, and United States for instance. This scheme is now broadly discussed for large scientific events. Assuming a minimum distance for flying of 700 km (and the mean of the \ademe, \myclimate, and \defra\ estimates), and assuming one hub in Europe, one in Japan and one in the United States, if each hub is located at the site associated with the minimum emission, the 100 tons could be reduced to $\sim 36$ tons of \co2eq\ by meeting in Geneva, Sagamihara, and Greenbelt, meaning a reduction by a factor of 3. This shall be considered as a maximum saving, as each hub meeting may receive attendees originally associated with another hub (e.g. youngest members of the Consortium, Consortium board members\ldots).
\item Locate always X-IFU related meetings at the site of minimum emission for the given audience, reachable by train within less than a day, close to international hubs which can be reached directly (to avoid connecting flights), with the list of attendees restricted to key persons, whose presence is mandatory. Changes in meeting organization may be needed to enable people to reach the meeting place by train, without impacting their private life, i.e. not having to travel over the weekends. 
\item Coordinate the presence of X-IFU to international conferences, restricting the number of attendees to the bare minimum, sending representatives of the project residing close to the conference sites, and privilege those conferences where there is a clear policy towards minimizing the travel footprint and/or which promote alternative ways of attendance (e.g., virtual participation). Give priority to the youngest consortium members to attend conferences, e.g., for presenting and promoting their work and contribution to the project but also for networking\ldots. Within X-IFU, one should accompany and support the transition to evaluating differently the evolution of academic careers, as to ensure that those who fly less are not penalized. 
\item Compensate: X-IFU is funded by public money. At the level of the funding agencies, there is no global compensation mechanism yet implemented. Compensation occurs only at personal level. Shall compensation be made possible in the future, its efficiency should however be demonstrated. In all cases, compensating should not be seen as the way to continue as before, but should come together with the undertaking of all efforts in reducing our emission in the first place.
\item Organize scientific conferences related to climate changes and its societal impacts as part of the plenary sessions of each consortium meetings. This should help in triggering a more global reflection on the way we keep our research programs going along the reduction path of our footprint, while at the same time supporting the development of similar projects in emerging countries through international cooperation.
\item Dedicate personnel to implement all the above actions in a coordinated way across the Consortium. 
\end{itemize}
Adopting a greener approach, whenever possible, will also be pursued, e.g., reducing waste, using re-usable items during X-IFU related events. Along the development of the X-IFU, many scientists and engineers will be exposed to the grand public. Each of us should engage with the public about the threat imposed by climate change and the impact of aviation, and should promote alternatives to flying to work collectively on a large scale project. 

This paper focuses entirely on the travel footprint of the project. Estimating the global footprint of the project is a humungous task which goes beyond the scope of this paper. Nevertheless, it will be computed and presented elsewhere.  

\section{Conclusions}
\label{conclusions}
A travel footprint calculator has been developed to monitor and reduce the travel footprint associated with the development of the Athena X-ray Integral Field Unit. The calculator uses seven different emission factors and methods leading to estimates that differ by up to a factor of $\sim 5$ for the same flying distance, calling for the need of standardization and regulation for estimating the aircraft emission and its associate impact on global warming. Those estimates should be based on a transparent, rigorous, reliable and indisputable methodology, worked out by an independent body. Amongst others, this is required for the science community to properly evaluate the impact of its activities, to question its contribution towards the goal of mitigating climate change in the best possible way, while reflecting globally on how to carry research in the decades to come, taking full benefits of the rapidly improving technologies of communication. 

Running the calculator on a set of X-IFU related meetings, the annual travel footprint of the project is about 500 tons. Actions are being implemented to lower this number, with strong incentive to use video conferences, whenever traveling is not absolutely required. Traveling will not be banned but each travel will be optimized. Reducing the travel footprint of the project will require the involvement of each consortium member and strong implication and exemplarity by the consortium leads, starting with the X-IFU PI. The initiative has received strong support throughout the X-IFU consortium, it now remains to convert this into daily acts. 

The full capabilities of the calculator have been illustrated by running it on the Fall meeting of the AGU and the lead author meetings of Working Group 1 of the IPCC. It has also been tested and used for some other applications (the X-ray Astronomy 2019 conference, the Cerenkov Telescope Array project, the EWASS 2019 conference,\ldots). It is released  for use by the science community at the following URL \url{https://travel-footprint-calculator.irap.omp.eu}.
\newpage
\begin{acknowledgements}
The author wishes to thank Vincent Albouys, the CNES X-IFU project manager, for his immediate support to the initiative of reducing the travel carbon footprint of the X-IFU project. I am grateful to Isabelle Dangeard (Institut Universitaire de Technologie de Quimper (IUT de Quimper)) for her help and guidance in understanding the various methods (and their shortcomings) in estimating carbon budgets and footprints, and for extensive comments to the manuscript. 

Very special thanks to Val\'erie Masson Delmotte, Co-Chair of IPCC Working Group I for discussions and for detailed and constructive comments all along the preparation of the paper. Thanks also to Jan Fuglestvedt, Vice Chair of IPCC WGI for insightful comments on the accounting of non-\co2\ effects. Thanks to Elisabeth Lonnoy for providing the list of participants of the 4 Lead Author Meetings of the IPCC Working Group I. 

Thanks to all the colleagues who have provided inputs along the development of the calculator, the website application and the comments on the paper: Leonard Burtscher, Deborah Costanzo, Felix Fuerst, Pierre-Louis Frison, Catherine Jeandel, Milan Kloewer, Jurgen Knoedelseder, Luigi Tibaldo, Maeva Voltz and St\'ephane Paltani. 

I would to thank Antoine Goutenoir for his enthusiasm, his dedication and skills in developing the web application of the calculator, and thank you to IRAP for hosting the web application of the calculator.

Finally, the author is grateful to an anonymous referee for a very careful reading of the paper.
\end{acknowledgements}
\bibliographystyle{spphys}
\bibliography{dbarret_vs}

\newpage

\section{Appendix A: Methodology used for the travel footprint calculator}\label{appendixa}
\subsection{How does the calculator work?}
A round trip is defined by a pair of cities. The two cities are geolocated and from their longitude and latitude the great circle distance is computed. This is the shortest path a plane can follow. Some methods thus consider uplift correction factors, to account for deviations from the shortest paths (due to holding patterns, avoidance of bad weather conditions\ldots)  in computing the emission of a flight (e.g. \defra\ uses a multiplying factor of 8\% to the emission coefficient, \myclimate\ uses instead a detour constant of 95 km independently of the distance flown, see \S\ref{myclimatemodel}). In addition, two cities may not be connected by a direct flight. Connecting flights are associated with larger emissions because they increase the travel distances, adding takeoffs/landings/\ldots which have a larger fuel consumption than during the cruise. This is accounted for by the calculator by increasing all great circle distances by 5\% (\cite{lhackel} considered 2.7\%, based on US airline statistics). Each method translates to a function giving the flight emission in kg as a function of the flight distance in km, see Figure \ref{dbarret_f1}. Thus from the increased great circle distance, the flight emission of a flight associated with a trip between a city pair is computed and multiplied by two to account for a round trip.

\subsection{Seating category}
\begin{figure}
\centerline{	\includegraphics[width=0.95\linewidth]{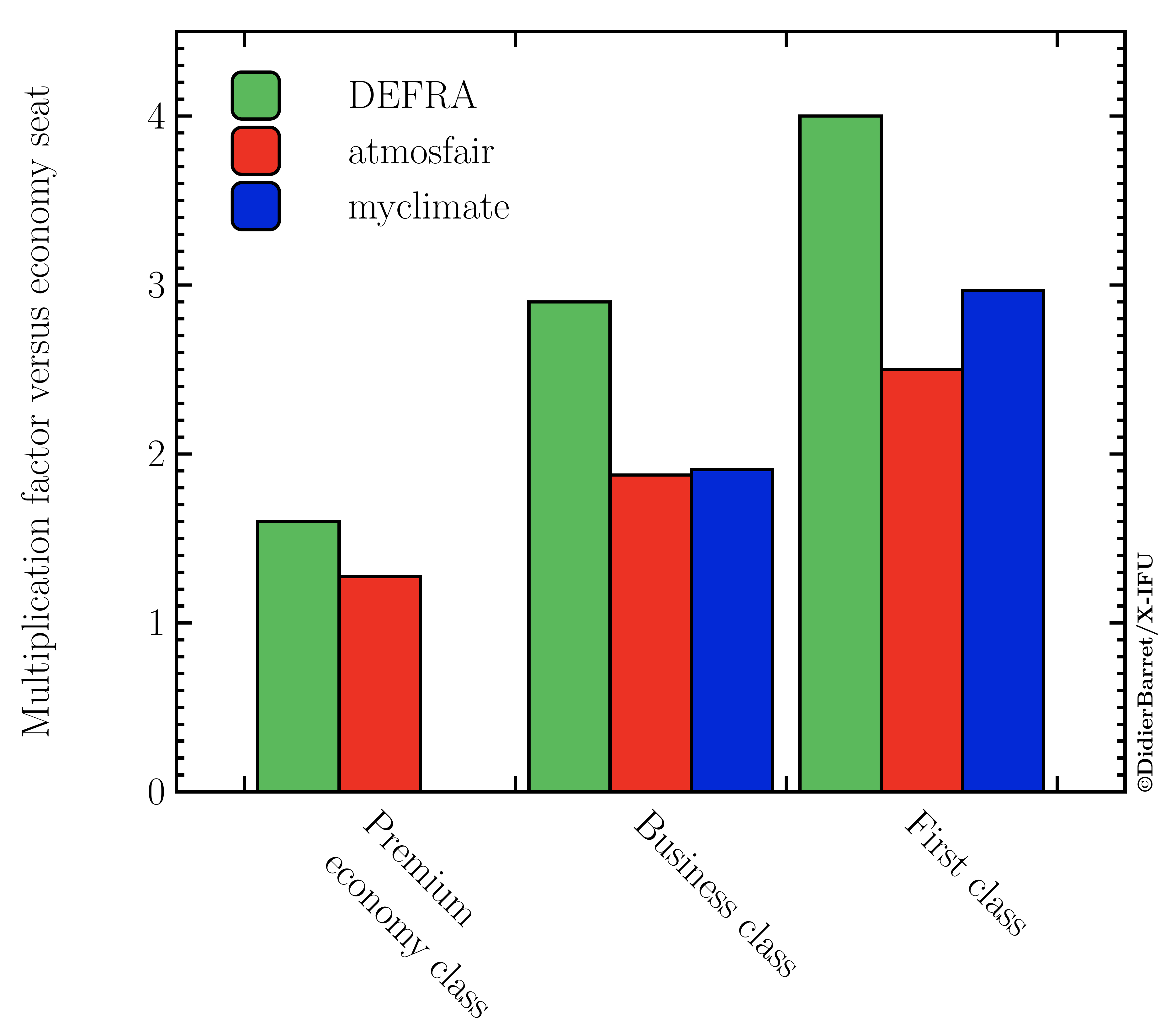} }
	\caption{Multiplication factor of the emission factors with respect to economy seats as considered by the \defra, \atmosfair, and \myclimate\ methods.} 
	\label{dbarret_f8} 
\end{figure}

The tool assumes economy seats for computing the flight emission. The emission factors for different seat classes relate to the area occupied by the seat in the plane. Figure \ref{dbarret_f8} shows the \defra, the \atmosfair, and the \myclimate\ multiplying factors inferred from their methodology. As can be seen, there are some differences between the different methods, but on average it can be considered that the footprint can be multiplied by $\sim 1.5$, $\sim2$ and $\sim 3$ for flying in Premium Economy, Business and First class. Note that the above factors should also depend in principle on the occupancy rate of the seats in each category.

\subsection{Approximating train travel distances from great circle distances}
\label{road_distance_vs_gcd}
\begin{figure}[!h]
	\centerline{\includegraphics[width=0.95\linewidth]{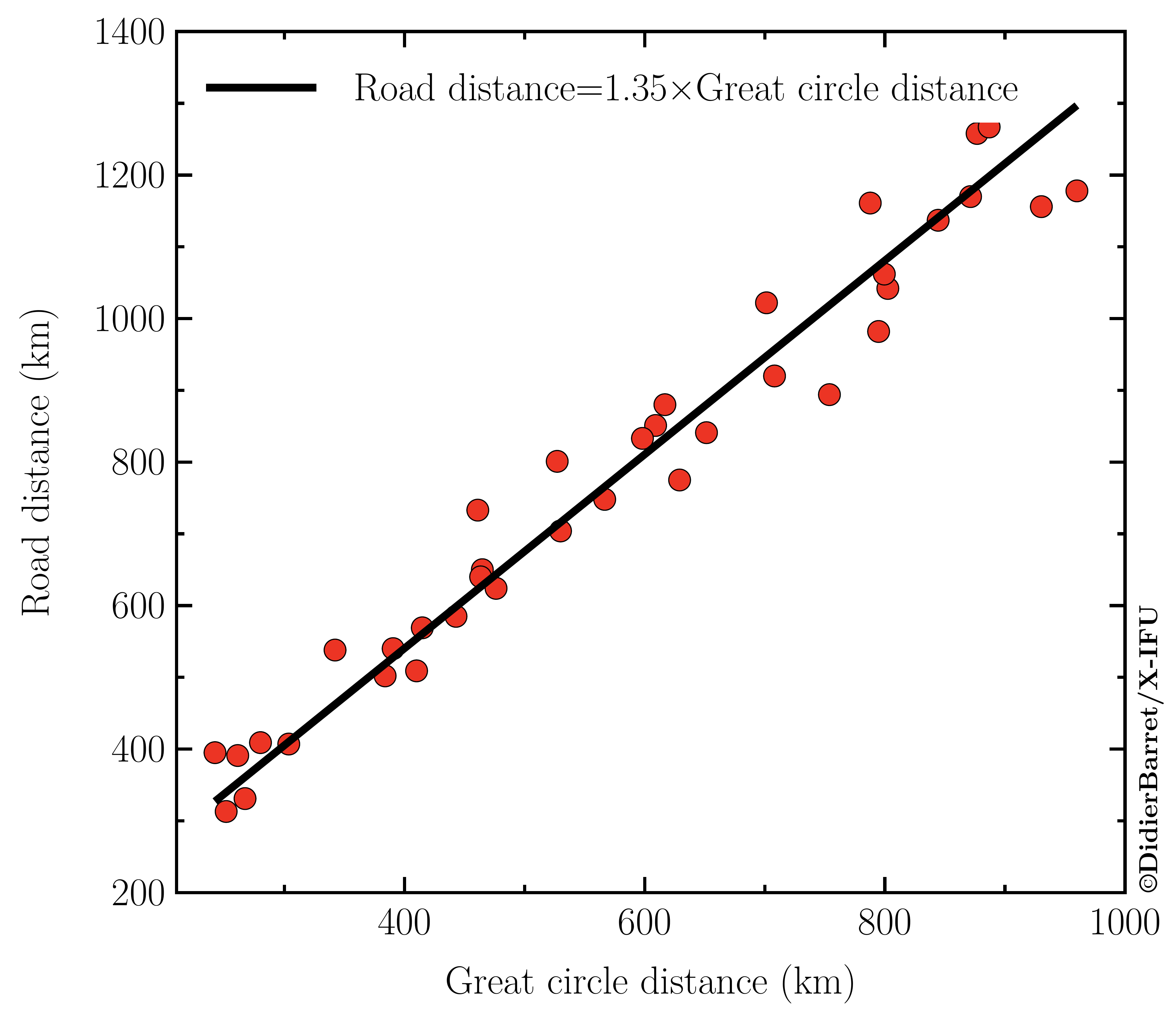}} 
	\caption{Road distances computed from {\it googlemaps} versus great circle distances computed from the X-IFU footprint calculator for a set of city pairs, with either Toulouse or Paris as the city of origin. The road distance is assumed to be a proxy of the train travel distance. Hence the train travel distance is computed from the great circle distance multiplied by 1.35.} 
	\label{dbarret_f9} 
\end{figure}

The footprint calculator enables to chose a train journey between city pairs, instead of a flight. To compute the train travel distance, one assumes that the shortest road distance between two cities is a proxy of the train travel distance. We have computed road distances from Toulouse and Paris to a set of cities across Europe, within $\sim 1300$ km. We plot road distances versus great circle distances in Figure \ref{dbarret_f9}. As can be seen, under the above assumption, with some scattering, road distances are found on average 35\% larger than  great circle distances. Since the calculator computes the great circle distances between city pairs, to compute the emission of the train journey, a 1.35 multiplication factor is applied to the emission associated with the measured great circle distance. 

Some may select train traveling instead of flying based on the duration of the trip. It is thus worth relating the minimum distance for flying to a travel time duration by train. In Europe, from a shallow survey of trip durations between major cities, it appears that the average speed of trains is about 100 km/hour (there are cases however for which excellent connections enables to reach speed significantly larger, up to 200 km/h). This means that on average it takes about 8 hours to travel by train between two cities with a great circle distance of 500 km (including connection, deviations from the shortest paths\ldots). 100 km/hour may be on the low side by a couple of 10 km/hour in countries where high speed trains are common, and on the high side by the same amount in countries where high speed trains are less available or not available at all.

\subsection{Input and output data for the calculator}
The web application of the X-IFU calculator is based on a form with very limited inputs requested. The inputs are provided in US English for the city and country names, without diacritics. On each line, the city and country names must be separated by a comma. Pasting a csv file in the form is possible, provided that a comma separates the city and country names. The calculator provides comma-separated-values (csv) and for convenience Excel (xlsx) template files, which can then be filled in and later uploaded. The first row of the files must be labeled with City and Country.

A round trip is defined by a city pair. If the user enters cityA as the origin city, and twice cityB as destinations, the tool returns the cumulative emission and distance from two round trips involving cityA and cityB, and indicates that 2 round trips were involved. The same happens if the user enters twice cityA as the city of origin and cityB as the sole destination.\\

The calculator can be run under three configurations: 

\begin{itemize}
\item A single city of origin and multiple destinations: this would be the case for an individual or an organization for which the city of departure is always the same. After grouping the city pairs, the calculator ranks the cities of destination according to the associated travel footprint.

\item Multiple cities of origin and a single destination: this would be the case for computing the footprint of a conference/meeting\ldots. After grouping the city pairs, the calculator ranks the cities of origin according to the associated travel footprint.

\item Multiple cities of origin and multiple destinations: this would be the case for finding the city corresponding to the minimum travel footprint or comparing the travel footprint associated with different hosts of an event. The site of minimum emission is obviously associated with the smallest summed travel distance when only flights are considered, but may differ slightly when train journeys are allowed. After grouping the city pairs, the calculator ranks the city of destination according to the associated travel footprint summed over all cities of origin. 

\end{itemize}

The user can select any combination of the seven methods implemented by the calculator. If more than one method is selected, the calculator returns the mean of the estimates of the selected methods.

The above cases can be run with the assumptions that all travels are by plane, or by considering a minimum distance for flying chosen between 100, 300, 500, 700 and 1000 km (500 km being the default option). 

The result page provides a summary plot which can be downloaded in svg and png formats (see Figure \ref{dbarret_f11} for a plot example), as well as a csv and raw yaml file, which can be used for further processing. The csv file lists the name of the city as in the form, the address to which it was geolocated, the \co2eq\ emission (in kg), the cumulative distance traveled, the number of trips possible by train (i.e. when the distance is less than the minimum flying distance, e.g. 500 km) and the number of trips by plane. The plot and the csv file always rank the cities against their associated footprint.

\subsection{Trouble shooting and caveats}
The estimation relies on the quality of the inputs provided. It can go wrong if a city is not properly geolocated. This may happen because the name of the city is wrongly spelled or the geolocator (OpenStreetMap, OSM) is confused. An error should be listed at the end of the result page. Do not be surprised, if the name recovered by the geolocator is not exactly the one you had expected (e.g. a city is located at the address of an embassy). An error may also occur if the input file submitted does not comply with the requested format (e.g. the first row is not properly labeled), including font encoding. If nothing happens during a request, it is most likely caused by the geolocator being unavailable. In this case, try again a few minutes later. The calculator has been run on some large input files, such as for the AGU meeting (24000 cities of origin and 20 possible destinations). The larger the input data set, the longer it takes and the higher the chances are that the geolocator gets stuck in the process. 

The numbers provided by the tool do not come with uncertainties, which must be pretty large as shown by the range of models considered in the calculator. In all cases however, the numbers can be used for relative comparisons, e.g. when comparing two cities for hosting a conference.

\newpage

\subsection{The X-IFU travel footprint calculator input form}
\begin{figure}[!h]
	\centerline{\includegraphics[width=0.85\linewidth]{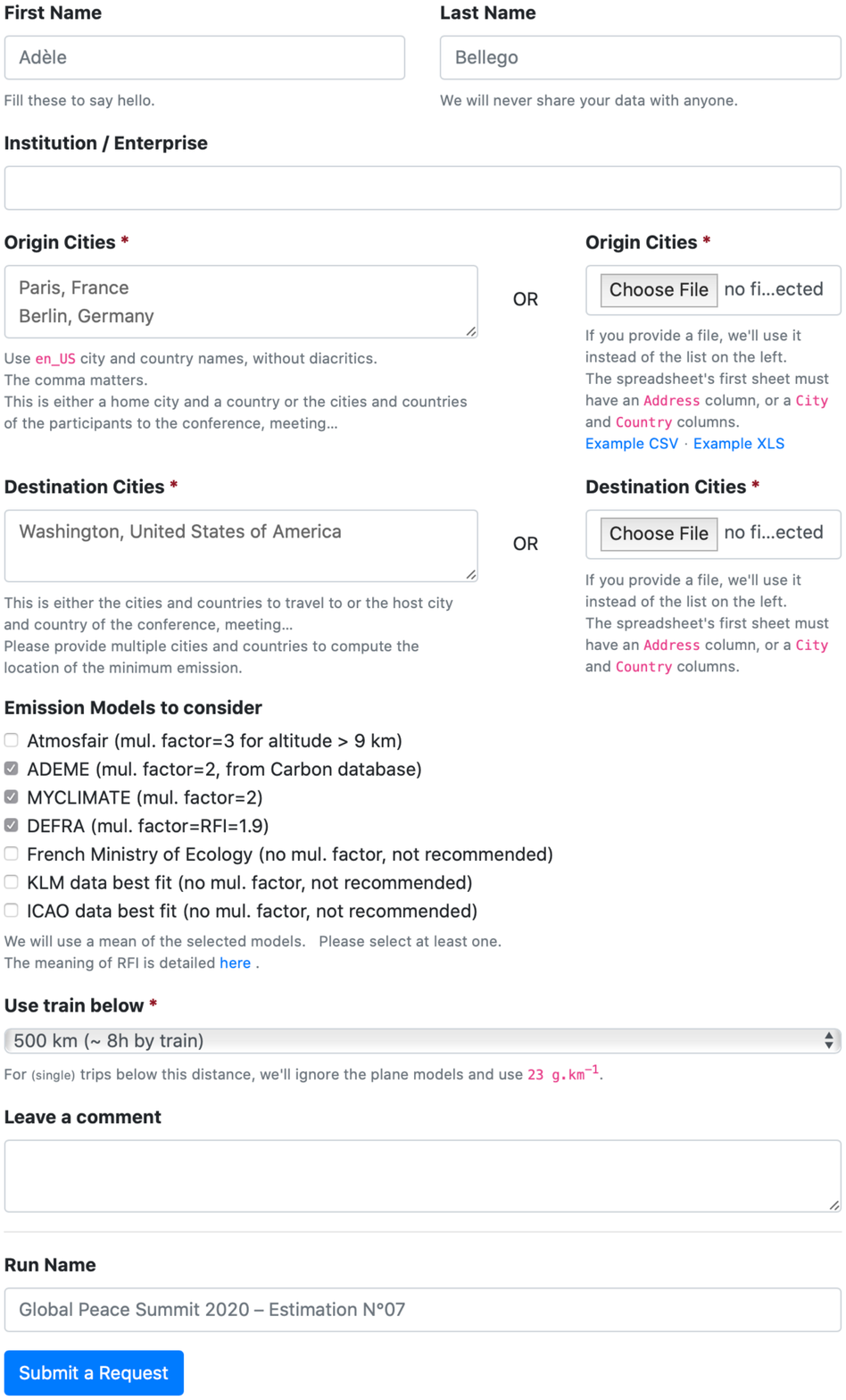}} 
	\caption{The form for entering the inputs to the calculator. } 
	\label{dbarret_f10} 
\end{figure}

Inputs to the X-IFU travel footprint calculator are to be provided through the form shown below (see Figure \ref{dbarret_f10}). The data provided through the form will remain confidential, as will be the results.

\newpage

\subsection{Graphical output of the travel footprint calculator}
\begin{figure*}[!htbp]
	\centerline{\includegraphics[clip,width=0.7\linewidth]{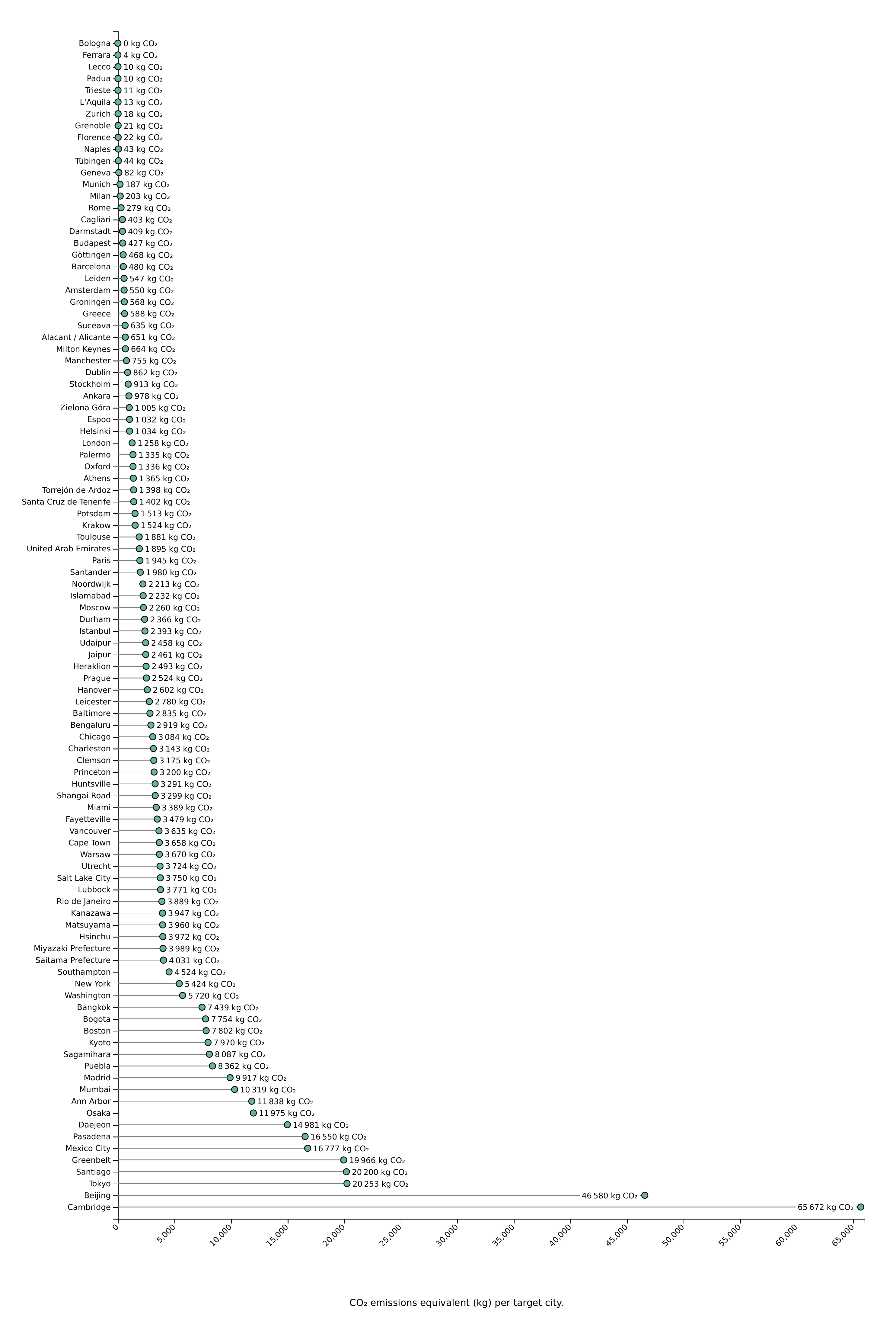}} 
	\caption{The graphical output of the online X-IFU travel footprint calculator run for the case of the X-ray Astronomy 2019.  The X-IFU calculator ranks the cities according to their associated travel footprint. The minimum distance for flying is assumed to be 500 km. Travels from cities at the top of the plot above Geneva (included) are assumed to be done by train. The \co2eq\ emission (summed over all the participants traveling from the given city) labels the green symbols on the plot (the number of travelers is not shown and the difference of emissions between two nearby cities is related to a different number travelers). The emission is computed from the mean of the \ademe, \atmosfair, \myclimate, and \defra\ estimates, assuming all passengers flew in economy class. The total emission from the meeting was $\sim 465$ tons of \co2eq, and for the audience of the meeting Bologna was close to the site of minimum emission. At the bottom of the figure, Cambridge refers to the city in the United States.} 
	\label{dbarret_f11} 
\end{figure*}

The X-ray Astronomy 2019 conference was held in Bologna, gathering about 340 participants (data courtesy of D. Costanzo). The graphical output of the footprint calculator is shown in Figure \ref{dbarret_f11}.

\newpage

\section{Appendix B: Emission factors from the different methods}\label{appendixb}
\subsection{Derived emission factors from \atmosfair, \icao, and KLM data}
\begin{figure}[!h]
	\centerline{\includegraphics[width=0.95\linewidth]{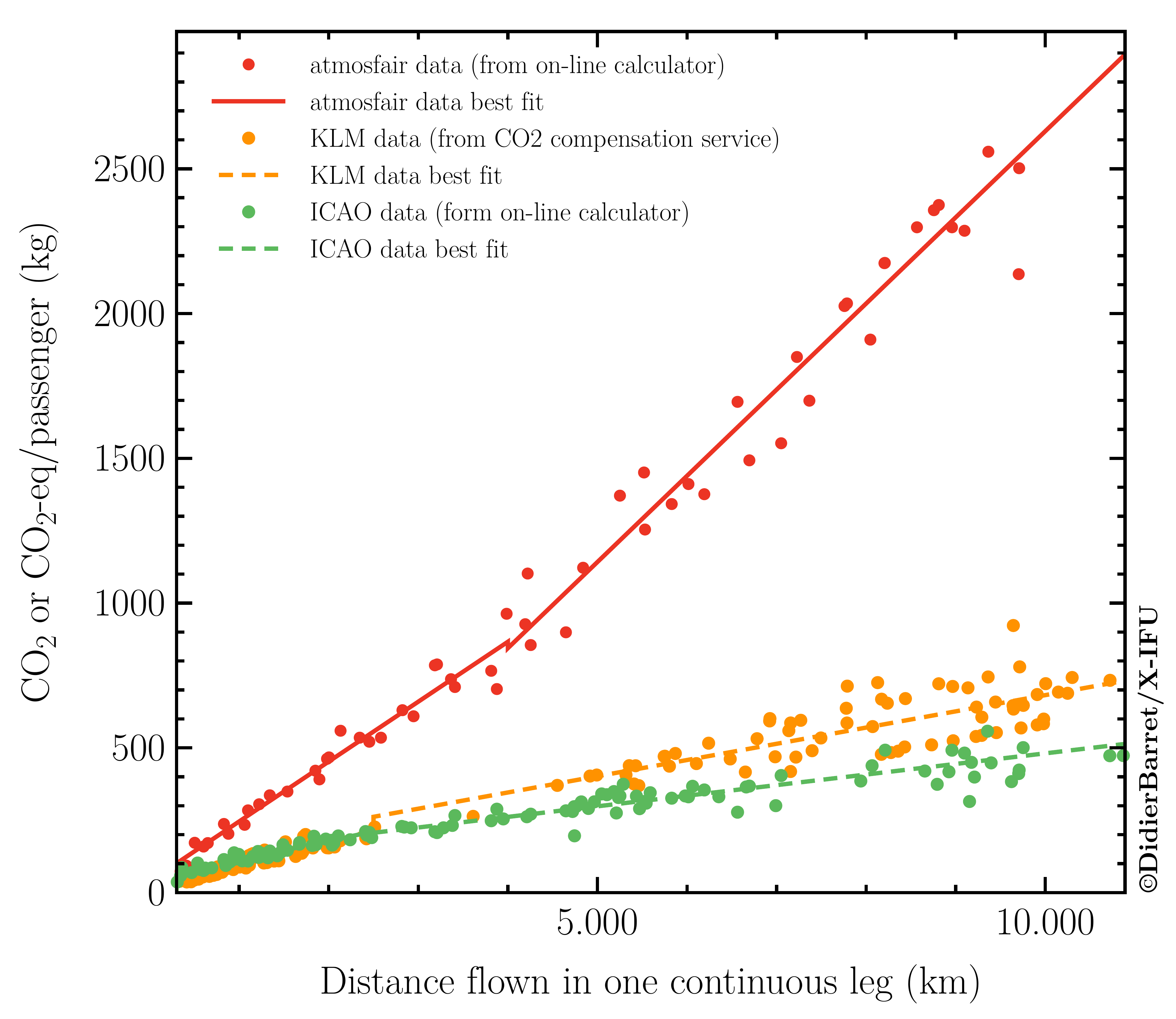}} 
	\caption{Two on-line calculators (\atmosfair\ and \icao) for which no analytical functions exist have been run for a wide range of distances and the provided estimates fitted by linear functions, within adjacent distance intervals. Data from the KLM compensation service web site have been retrieved and fitted the same way. The differences between the \atmosfair\ and the other two methods are explained only in part by the fact that \icao\ and KLM do not consider a multiplication factor for non-\co2\ effects, while \atmosfair\ considers a multiplication factor of 3 for altitudes larger than 9 km. } 
	\label{dbarret_f12} 
\end{figure}

The on-line calculators have been run for a range of flight distances for \atmosfair, and \icao\ \cite{atmosfair,icao}. The KLM data have been extracted from their Carbon compensation service website. The data have then been fitted by linear functions. The data points and the best fits are shown in Figure \ref{dbarret_f12}. Deviations of up to a few tens of \% can be found between the data points and the best fits. \icao\ does not include a multiplier for non-\co2\ effects in its estimates. No information is provided by KLM on that issue, but given the proximity of the estimates with the \icao\ ones it seems very likely that KLM also ignores radiative forcing in the estimates provided on their website.

\begin{figure}[!h]
\begin{center}
\includegraphics[width=0.95\hsize]{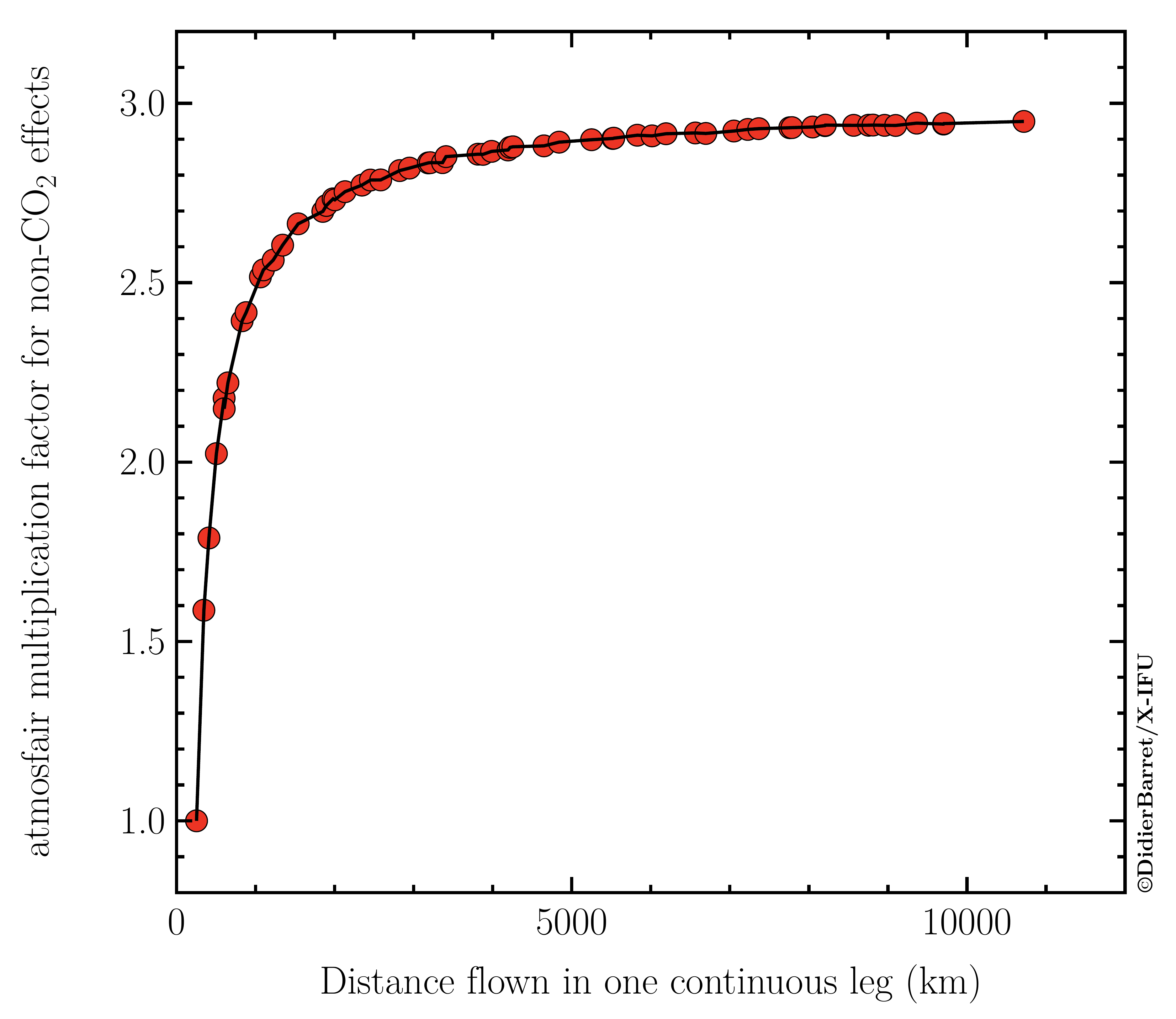}
\caption{The \atmosfair\ carbon emission calculator provides two estimates for the flight emission: one value which corresponds to the estimate of the fuel burnt and one value which provides the other part of the climate impact (contrails, ozone formation and other effects). This plots shows the ratio between the latter and the former. A multiplier coefficient of 3 is applied for all emissions above 9 km, which translates to a multiplication factor of 2.95 for long distance flights.}
\label{dbarret_f13}
\end{center}
\end{figure}

The direct \co2\ emission multiplier used by \atmosfair\ is plotted in figure \ref{dbarret_f13} as a function of the great circle distance. Accounting for the profile of the flight, the multiplier reaches around $\sim 2.95$ for long distance flights. 

\subsection{Mean emission factors from the French \ademe\ Carbon database}
\ademe\ provides mean emission factors as a function of leg distance and seat capacity of the plane \cite{ademe}. For the X-IFU calculator, for each leg, we have assumed the emission factors as the average over all seat capacities of the plane, as listed in the last column of Table \ref{tab:basecarbone}. It is worth noting that \ademe\  provides emission factors for planes of low seat capacity, and that those are significantly larger than the ones associated with bigger planes. Such small high emitting planes may not be part of the fleet considered by the other calculators. The same applies to the data provided by the French Ministry for the Ecological and Inclusive Transition (\S\ref{ministry}).
\begin{table}[!htbp]
   \centering

   \begin{tabular}{lcccccc} 
   \hline 
Distance & $<50$  & 50-100  & 100-180  & 180-250  & $>250$ & Average\\
 (km) & seats & seats & seats & seats & seats  \\
    \hline 
0-1000  & 0.683 & 0.453 & 0.314 & 0.293 & \ldots     & 0.4358 \\ 
1000-2000  & 0.906 & 0.314 & 0.258 & 0.216 &\ldots    & 0.4235 \\ 
2000-3000  & 1.2 & 0.209 & 0.237 & 0.209 & \ldots   & 0.4638 \\ 
3000-4000  &  \ldots   & \ldots    & 0.230 & 0.230 & 0.251 & 0.2370 \\ 
4000-5000  &  \ldots  &  \ldots   & 0.293 & 0.307 & 0.258 & 0.2860 \\ 
5000-6000  &   \ldots  &  \ldots   & 0.286 & 0.230 & 0.223 & 0.2463 \\
6000-7000  &  \ldots   &  \ldots   &  \ldots   & 0.223 & 0.209 & 0.2160 \\ 
7000-8000  &   \ldots  & \ldots   & \ldots   & 0.202 & 0.209 & 0.2055 \\ 
8000-9000  &  \ldots   &   \ldots  &  \ldots   & 0.223 & 0.23 & 0.2265 \\
9000-10000  &   \ldots  &   \ldots  & \ldots    & 0.216 & 0.223 & 0.2195 \\
10000-11000  &  \ldots  &  \ldots   &  \ldots   &  \ldots   & 0.216 & 0.2160 \\
$>$11000 km  &  \ldots  &  \ldots   &  \ldots   &  \ldots   & 0.223 & 0.2230 \\
\hline
\end{tabular}
    \caption{\ademe\ Carbon database emission coefficient for flights in kg of \co2eq\ per km.passenger \cite{ademe}. The data can be retrieved from the URL: \url{http://www.bilans-ges.ademe.fr}, but requires identification.}
   \label{tab:basecarbone}
\end{table}
\subsection{\defra}
\defra\ provides mean emission factors for different flight types (see Table \ref{tab:defra}). The values include a 1.9 multiplier to account for non-\co2\ effects. The multiplier is referred to as a Radiative Forcing Index by \defra. Short haul are for flights below 3700 km and long-haul above. For computing the \defra\ emission of long haul flights, the mean of the emission coefficients of flights from/to the UK and international flights has been assumed. 

 \begin{table}[!htp]
\begin{center}
 \begin{tabular}{lc} 
   \hline 
Haul & kg \co2eq\ \\
    \hline
  Short-haul, to/from UK &  0.156 \\
  Long-haul, to/from UK &  0.150\\
  International, to/from non-UK &  0.138   \\
  \hline
    \end{tabular}
    \caption{The \defra\ emission factors taken from \cite{defra} in kg of \co2\ per passenger.km. A 1.9 multiplier is included in the emission factors.}
   \label{tab:defra}
   \end{center}
\end{table}
\subsection{Mean emission factors from the French Ministry for the Ecological and Inclusive Transition.}\label{ministry}
The  French Ministry for the Ecological and Inclusive Transition provides mean emission factors as a function of leg distance and seat capacity of the plane, in a similar way as \ademe\ \cite{ministryecology}. For the calculator, we have averaged the emission factors over all seat capacities of the plane, as listed in the last column of Table \ref{tab:ministry}. No radiative forcing is included in these emission coefficients. Similar to \ademe\, the ministry provides high emission factors for small capacity planes.

\begin{table}[!htbp]
   \centering
    \caption{Mean emission coefficients as derived from green house gas information for transport services from the Ministry for the Ecological and Inclusive Transition. These are the emission coefficients for flights in kg of \co2\ per passenger.km. Reference: GHG information for transport services. Application of Article L. 1431-3 of the French transport code. Methodological guide. Updated version resulting 67 article of the law n$^\circ$ 2015-992. Version of June 2019. Taken from Table 5 and 21 bis from \cite{ministryecology}.}
   \begin{tabular}{lcccccc} 
   \hline 
Distance & $<50$  & 50-100  & 101-180  & 181-250  & $>250$ & Average \\
 (km) & seats & seats & seats & seats & seats & \ldots \\
    \hline 
0-1000  & 0.223 & 0.187 & 0.141 & 0.117 & \ldots  & 0.167\\
1000-2000  & 0.254 & 0.161 & 0.117 & 0.095 & 0.123 & 0.150\\
2000-3000  & \ldots  & \ldots  & 0.109 & 0.091 & 0.101 & 0.100\\
3000-4000  & \ldots  & \ldots  & 0.105 & 0.099 &0.099 &0.101\\
4000-5000  & \ldots  & \ldots  & 0.153 & 0.126 & 0.090 &0.123\\
5000-6000  & \ldots  & \ldots  & 0.150 & 0.098 & 0.088 &0.112\\
6000-7000  & \ldots  & \ldots  & \ldots  & 0.100 & 0.082 &0.091\\
7000-8000  & \ldots  & \ldots  & \ldots  & 0.091 & 0.087 &0.089\\
8000-9000  & \ldots  & \ldots  & \ldots  & 0.095 & 0.087 &0.091\\
9000-10000  & \ldots  & \ldots  & \ldots  & 0.073 & 0.083 &0.078\\
10000-11000  & \ldots  & \ldots  & \ldots  & \ldots  & 0.095 &0.095\\
$>$11000 km  & \ldots  & \ldots  & \ldots  & \ldots  & 0.094 &0.094\\
   \hline 
   \end{tabular}
   \label{tab:ministry}
\end{table}
\subsection{\myclimate}\label{myclimatemodel}
\myclimate\ provides an analytical formula as given below \cite{myclimate}, whose parameters are listed in Table \ref{tab:myclimate}.: 
\begin{equation}
E=\frac{ a x^2 +b x +c }{S \times PLF} \times (1-CF) \times CW \times (EF \times M + P) + AF \times x + A 
\end{equation}

in which 
\begin{itemize}
\item E: \co2-eq emissions per passenger [kg]
\item x: Flight Distance [km] which is defined as the sum of GCD, the great circle distance, and DC, a distance correction for detours and holding patterns, and inefficiencies in the air traffic control systems [km]
\item S: Average number of seats (total across all cabin classes)
\item PLF: Passenger load factor
\item CF: Cargo factor
\item CW: Cabin class weighting factor
\item EF: \co2\ emission factor for jet fuel combustion (kerosene)
\item M: Multiplier accounting for potential non-\co2\ effects
\item P: \co2eq\ emission factor for preproduction jet fuel, kerosene
\item AF: Aircraft factor
\item A: Airport infrastructure emissions
\item The part $a x^2 + bx + c$ is a nonlinear approximation of f(x) + LTO
\item LTO: Fuel consumption during landing and takeoff cycle including taxying [kg]
\item Short-haul is defined as $x<1500$ km and long-haul as $x>2500$ km. In between, a linear interpolation is used.
\end{itemize}

\begin{table}[!htbp]
   \centering
   \begin{tabular}{lcc} 
   \hline 
 Aircraft type & Generic short-haul & Generic long-haul \\
 \hline
Average seat number (S) & 153.51& 280.21\\
Passenger load factor (PLF)& 0.82 & 0.82 \\
Detour constant (DC)&95 &95 \\
1 - Cargo factor (1 - CF) & 0.93& 0.74\\
Economy class (CW) & 0.96& 0.80 \\
Business class weight (CW) & 1.26 & 1.54 \\
First class weight (CW) & 2.40& 2.40 \\
Emission factor (EF) & 3.15& 3.15\\
Preproduction (P)& 0.54& 0.54\\
Multiplier (M) & 2& 2\\
Aircraft factor (AF) & 0.00038& 0.00038\\
Airport/Infrastructure (A)& 11.68& 11.68\\
a& 0. & 0.0001 \\
b& 2.714 & 7.104 \\
c& 1166.52 & 5044.93 \\
   \hline
   \end{tabular}
    \caption{\myclimate\ model parameters as taken from \cite{myclimate}.}  
   \label{tab:myclimate}
\end{table}
\subsection{Normalization of the emission coefficients}
\begin{figure}[!ht]
\centerline{\includegraphics[width=0.95\linewidth]{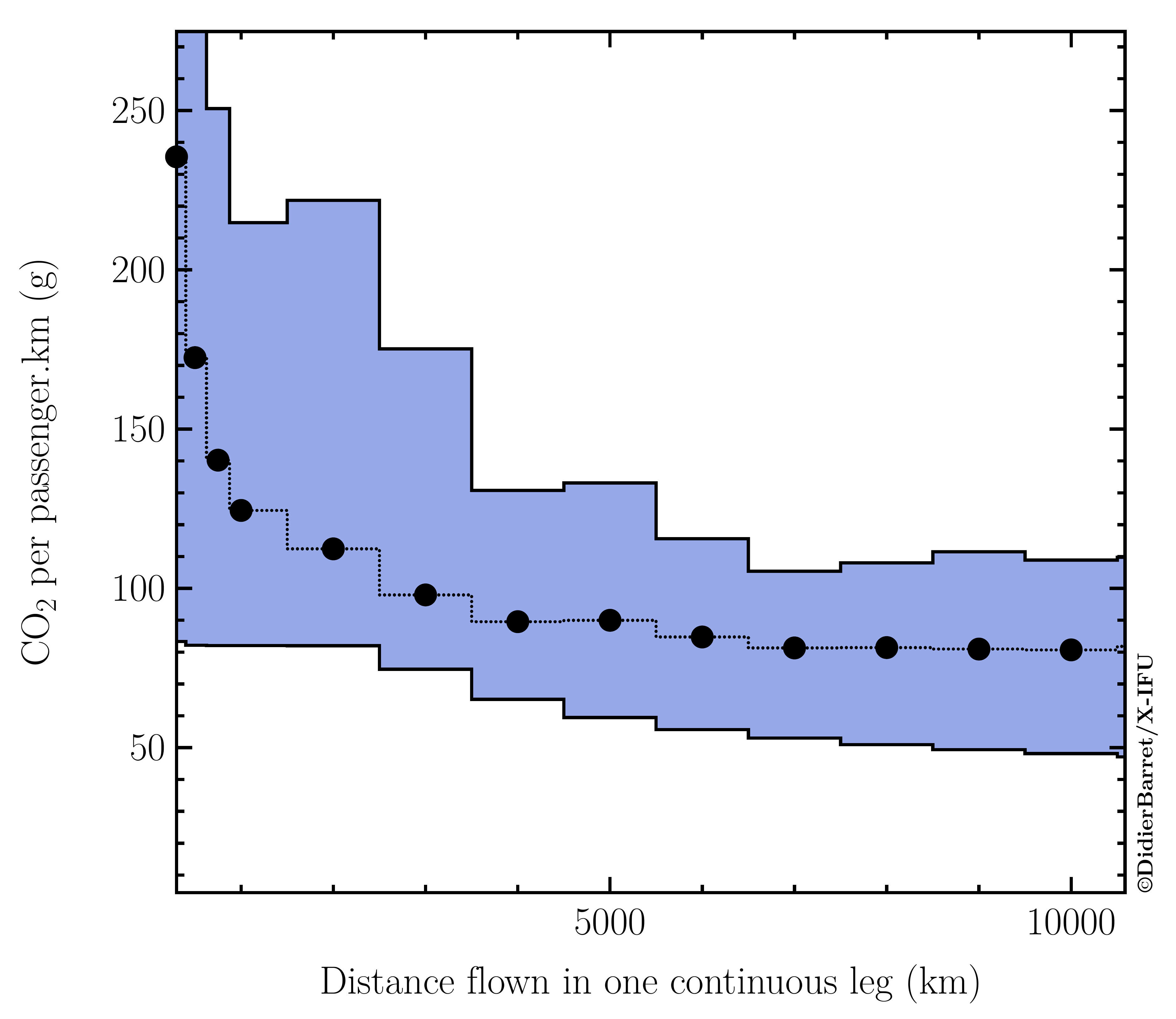} }
	\caption{Range of emission factors (\co2\ per passenger.km in g) as a function of the distance flown in one continuous leg (km). The lower and upper boundaries of the colored region are given by the minimum and maximum emission factors of the seven methods used by the calculator. The mean values of emission factors of the seven methods are shown with filled black circles. No multiplication factor, accounting for non-\co2\ effect is applied, enabling the methods to be compared. The emission is computed assuming economy seating. } 
	\label{dbarret_f14} 
\end{figure}

It is interesting to compare the emission coefficients of the different methods without any multiplier, accounting for non-\co2\ effects. Figure \ref{dbarret_f14} shows the range of the emission coefficients derived from the seven estimators. As can be seen, for short flying distances, a factor of $\sim 3$ is observed between the minimum and maximum of the emission factors. The factor goes to $\sim 2$ for large flying distances. For those distances, the mean emission factor levels off around $\sim 80 \pm 30 $ g of \co2\ per passenger.km. Although not apparent in the figure, the emission coefficients should reflect the fact that aircrafts are generally most efficient at intermediate distances, as for short distances the energy-intensive take-off and climb sections represent a large fraction of the fuel consumed, while for long distance flights the fuel to be carried at take-off and climb-out requires extra energy. 

\subsection{\defra\ train emission factors}
\label{defra_train_emission_factors}
For \defra\, the international rail factor ($\sim 6$ g of \co2 per passenger.km) is based on a passenger-km weighted average of the emission factors for Eurostars routes, e.g., London-Brussels, London-Paris. The methodology applied in calculating the Eurostar emission factors accounts for the total electricity use by Eurostar trains on the UK and France/Belgium track sections, the total passenger numbers (and therefore calculated passenger km) on all Eurostar services, the emission factors for electricity (in kg\co2\ per kWh) for the UK and France/Belgium journey sections. On the other hand, the national rail factor refers to an average emission per passenger km for diesel and electric trains in 2017-18 ($\sim 40$ g of \co2 per passenger.km). For both international and national rail factors, the CH$_4$ and N$_2$O emission factors have been estimated from the corresponding emission factors for electricity generation, proportional to the \co2\ emission factors. The sum of \co2\, CH$_4$ and N$_2$O emissions is expressed in \co2eq\ (see \cite{defra} for more details).

\end{document}